\documentclass[twocolumn,appendixfloats,twocolappendix]{aastex63}

\usepackage{natbib}
\usepackage{amsmath,amssymb}
\usepackage{xspace}
\usepackage{rotating}
\usepackage{multirow}
\usepackage{enumitem}

\definecolor{mpl_blue}{HTML}{1F77B4}
\definecolor{mpl_orange}{HTML}{FF7F0E}
\definecolor{mpl_green}{HTML}{2CA02C}
\definecolor{mpl_red}{HTML}{D62728}

\usepackage[all]{hypcap}
\usepackage[caption=false]{subfig}
\newcommand{\tempo}{{\sc tempo}\xspace}
\newcommand{\tempotwo}{{\sc tempo2}\xspace}
\newcommand{\pint}{{\sc pint}\xspace}

\newcommand{\bayesephem}{\textsc{BayesEphem}}

\newcommand{\m}[1]{\mathbf{#1}}

\def\be{\begin{equation}}
\def\ee{\end{equation}}
\newcommand{\bb}{\begin{bmatrix}}
\newcommand{\eb}{\end{bmatrix}}
\def\bea{\begin{eqnarray}}
\def\eea{\end{eqnarray}}

\def\R{\color{red}}

\newcommand{\Agw}{\ensuremath{A_\mathrm{GWB}}}
\newcommand{\Acp}{\ensuremath{A_\mathrm{CP}}}

\newcommand{\fyr}{f_\mathrm{yr}}

\defcitealias{abb+15}{NG9}
\defcitealias{abb+16}{NG9gwb}
\defcitealias{abb+18a}{NG11}
\defcitealias{abb+18b}{NG11gwb}
\defcitealias{aab+20}{NG12}

\begin{document}

\title{The NANOGrav 12.5-year Data Set: \\ Search For An Isotropic Stochastic Gravitational-Wave Background }
\shorttitle{NANOGrav 12.5-year Gravitational-Wave Background}
\shortauthors{The NANOGrav Collaboration}

% DO NOT EDIT THIS FILE. EDITS WILL BE OVERWRITTEN.
% AUTO-GENERATED WITH make-aastex62-author-list.py
% FROM authorship/author_list_12yr_data.txt, authorship/author_affil_and_orcid.txt, AND authorship/affil.txt
\author{Zaven Arzoumanian}
\affiliation{X-Ray Astrophysics Laboratory, NASA Goddard Space Flight Center, Code 662, Greenbelt, MD 20771, USA}
\author[0000-0003-2745-753X]{Paul T. Baker}
\affiliation{Department of Physics and Astronomy, Widener University, One University Place, Chester, PA 19013, USA}
\author[0000-0003-4046-884X]{Harsha Blumer}
\affiliation{Department of Physics and Astronomy, West Virginia University, P.O. Box 6315, Morgantown, WV 26506, USA}
\affiliation{Center for Gravitational Waves and Cosmology, West Virginia University, Chestnut Ridge Research Building, Morgantown, WV 26505, USA}
\author[0000-0003-0909-5563]{Bence B\'{e}csy}
\affiliation{Department of Physics, Montana State University, Bozeman, MT 59717, USA}
\author[0000-0001-6341-7178]{Adam Brazier}
\affiliation{Cornell Center for Astrophysics and Planetary Science and Department of Astronomy, Cornell University, Ithaca, NY 14853, USA}
\affiliation{Cornell Center for Advanced Computing, Cornell University, Ithaca, NY 14853, USA}
\author[0000-0003-3053-6538]{Paul R. Brook}
\affiliation{Department of Physics and Astronomy, West Virginia University, P.O. Box 6315, Morgantown, WV 26506, USA}
\affiliation{Center for Gravitational Waves and Cosmology, West Virginia University, Chestnut Ridge Research Building, Morgantown, WV 26505, USA}
\author[0000-0003-4052-7838]{Sarah Burke-Spolaor}
\affiliation{Department of Physics and Astronomy, West Virginia University, P.O. Box 6315, Morgantown, WV 26506, USA}
\affiliation{Center for Gravitational Waves and Cosmology, West Virginia University, Chestnut Ridge Research Building, Morgantown, WV 26505, USA}
\affiliation{CIFAR Azrieli Global Scholars program, CIFAR, Toronto, Canada}
\author[0000-0002-2878-1502]{Shami Chatterjee}
\affiliation{Cornell Center for Astrophysics and Planetary Science and Department of Astronomy, Cornell University, Ithaca, NY 14853, USA}
\author[0000-0002-3118-5963]{Siyuan Chen}
\affiliation{Station de Radioastronomie de Nancay, Observatoire de Paris, Universite PSL, CNRS, Universite d'Orleans, 18330 Nancay, France}
\affiliation{FEMTO-ST Institut de recherche, Department of Time and Frequency, UBFC and CNRS, ENSMM, 25030 Besancon, France}
\affiliation{Laboratoire de Physique et Chimie de l'Environment et de l'Espace, LPC2E UMR7328, Universite d'Orleans, CNRS, 45071 Orleans, France}
\author[0000-0002-4049-1882]{James M. Cordes}
\affiliation{Cornell Center for Astrophysics and Planetary Science and Department of Astronomy, Cornell University, Ithaca, NY 14853, USA}
\author[0000-0002-7435-0869]{Neil J. Cornish}
\affiliation{Department of Physics, Montana State University, Bozeman, MT 59717, USA}
\author[0000-0002-2578-0360]{Fronefield Crawford}
\affiliation{Department of Physics and Astronomy, Franklin \& Marshall College, P.O. Box 3003, Lancaster, PA 17604, USA}
\author[0000-0002-6039-692X]{H. Thankful Cromartie}
\affiliation{Cornell Center for Astrophysics and Planetary Science and Department of Astronomy, Cornell University, Ithaca, NY 14853, USA}
\author[0000-0002-2185-1790]{Megan E. DeCesar}
\altaffiliation{NANOGrav Physics Frontiers Center Postdoctoral Fellow}
\affiliation{Department of Physics, Lafayette College, Easton, PA 18042, USA}
\affiliation{George Mason University, Fairfax, VA 22030, resident at U.S. Naval Research Laboratory, Washington, D.C. 20375, USA}
\author[0000-0002-6664-965X]{Paul B. Demorest}
\affiliation{National Radio Astronomy Observatory, 1003 Lopezville Rd., Socorro, NM 87801, USA}
\author[0000-0001-8885-6388]{Timothy Dolch}
\affiliation{Department of Physics, Hillsdale College, 33 E. College Street, Hillsdale, Michigan 49242, USA}
\author{Justin A. Ellis}
\affiliation{Infinia ML, 202 Rigsbee Avenue, Durham NC, 27701}
\author{Elizabeth C. Ferrara}
\affiliation{NASA Goddard Space Flight Center, Greenbelt, MD 20771, USA}
\author[0000-0001-5645-5336]{William Fiore}
\affiliation{Department of Physics and Astronomy, West Virginia University, P.O. Box 6315, Morgantown, WV 26506, USA}
\affiliation{Center for Gravitational Waves and Cosmology, West Virginia University, Chestnut Ridge Research Building, Morgantown, WV 26505, USA}
\author[0000-0001-8384-5049]{Emmanuel Fonseca}
\affiliation{Department of Physics, McGill University, 3600  University St., Montreal, QC H3A 2T8, Canada}
\author[0000-0001-6166-9646]{Nathan Garver-Daniels}
\affiliation{Department of Physics and Astronomy, West Virginia University, P.O. Box 6315, Morgantown, WV 26506, USA}
\affiliation{Center for Gravitational Waves and Cosmology, West Virginia University, Chestnut Ridge Research Building, Morgantown, WV 26505, USA}
\author[0000-0001-8158-638X]{Peter A. Gentile}
\affiliation{Department of Physics and Astronomy, West Virginia University, P.O. Box 6315, Morgantown, WV 26506, USA}
\affiliation{Center for Gravitational Waves and Cosmology, West Virginia University, Chestnut Ridge Research Building, Morgantown, WV 26505, USA}
\author[0000-0003-1884-348X]{Deborah C. Good}
\affiliation{Department of Physics and Astronomy, University of British Columbia, 6224 Agricultural Road, Vancouver, BC V6T 1Z1, Canada}
\author[0000-0003-2742-3321]{Jeffrey S. Hazboun}
\altaffiliation{NANOGrav Physics Frontiers Center Postdoctoral Fellow}
\affiliation{University of Washington Bothell, 18115 Campus Way NE, Bothell, WA 98011, USA}
\author{A. Miguel Holgado}
\affiliation{Department of Astronomy and National Center for Supercomputing Applications, University of Illinois at Urbana-Champaign, Urbana, IL 61801, USA}
\affiliation{McWilliams Center for Cosmology and Department of Physics, Carnegie Mellon University, Pittsburgh PA, 15213, USA}
\author{Kristina Islo}
\affiliation{Center for Gravitation, Cosmology and Astrophysics, Department of Physics, University of Wisconsin-Milwaukee,\\ P.O. Box 413, Milwaukee, WI 53201, USA}
\author[0000-0003-1082-2342]{Ross J. Jennings}
\affiliation{Cornell Center for Astrophysics and Planetary Science and Department of Astronomy, Cornell University, Ithaca, NY 14853, USA}
\author[0000-0001-6607-3710]{Megan L. Jones}
\affiliation{Center for Gravitation, Cosmology and Astrophysics, Department of Physics, University of Wisconsin-Milwaukee,\\ P.O. Box 413, Milwaukee, WI 53201, USA}
\author[0000-0002-3654-980X]{Andrew R. Kaiser}
\affiliation{Department of Physics and Astronomy, West Virginia University, P.O. Box 6315, Morgantown, WV 26506, USA}
\affiliation{Center for Gravitational Waves and Cosmology, West Virginia University, Chestnut Ridge Research Building, Morgantown, WV 26505, USA}
\author[0000-0001-6295-2881]{David L. Kaplan}
\affiliation{Center for Gravitation, Cosmology and Astrophysics, Department of Physics, University of Wisconsin-Milwaukee,\\ P.O. Box 413, Milwaukee, WI 53201, USA}
\author[0000-0002-6625-6450]{Luke Zoltan Kelley}
\affiliation{Center for Interdisciplinary Exploration and Research in Astrophysics (CIERA), Northwestern University, Evanston, IL 60208}
\author[0000-0003-0123-7600]{Joey Shapiro Key}
\affiliation{University of Washington Bothell, 18115 Campus Way NE, Bothell, WA 98011, USA}
\author{Nima Laal}
\affiliation{Department of Physics, Oregon State University, Corvallis, OR 97331, USA}
\author[0000-0003-0721-651X]{Michael T. Lam}
\affiliation{School of Physics and Astronomy, Rochester Institute of Technology, Rochester, NY 14623, USA}
\affiliation{Laboratory for Multiwavelength Astrophysics, Rochester Institute of Technology, Rochester, NY 14623, USA}
\author{T. Joseph W. Lazio}
\affiliation{Jet Propulsion Laboratory, California Institute of Technology, 4800 Oak Grove Drive, Pasadena, CA 91109, USA}
\author[0000-0003-1301-966X]{Duncan R. Lorimer}
\affiliation{Department of Physics and Astronomy, West Virginia University, P.O. Box 6315, Morgantown, WV 26506, USA}
\affiliation{Center for Gravitational Waves and Cosmology, West Virginia University, Chestnut Ridge Research Building, Morgantown, WV 26505, USA}
\author{Jing Luo}
\affiliation{Department of Astronomy \& Astrophysics, University of Toronto, 50 Saint George Street, Toronto, ON M5S 3H4, Canada}
\author[0000-0001-5229-7430]{Ryan S. Lynch}
\affiliation{Green Bank Observatory, P.O. Box 2, Green Bank, WV 24944, USA}
\author[0000-0003-2285-0404]{Dustin R. Madison}
\altaffiliation{NANOGrav Physics Frontiers Center Postdoctoral Fellow}
\affiliation{Department of Physics and Astronomy, West Virginia University, P.O. Box 6315, Morgantown, WV 26506, USA}
\affiliation{Center for Gravitational Waves and Cosmology, West Virginia University, Chestnut Ridge Research Building, Morgantown, WV 26505, USA}
\author[0000-0001-7697-7422]{Maura A. McLaughlin}
\affiliation{Department of Physics and Astronomy, West Virginia University, P.O. Box 6315, Morgantown, WV 26506, USA}
\affiliation{Center for Gravitational Waves and Cosmology, West Virginia University, Chestnut Ridge Research Building, Morgantown, WV 26505, USA}
\author[0000-0002-4307-1322]{Chiara M. F. Mingarelli}
\affiliation{Center for Computational Astrophysics, Flatiron Institute, 162 5th Avenue, New York, New York, 10010, USA}
\affiliation{Department of Physics, University of Connecticut, 196 Auditorium Road, U-3046, Storrs, CT 06269-3046, USA}
\author[0000-0002-3616-5160]{Cherry Ng}
\affiliation{Dunlap Institute for Astronomy and Astrophysics, University of Toronto, 50 St. George St., Toronto, ON M5S 3H4, Canada}
\author[0000-0002-6709-2566]{David J. Nice}
\affiliation{Department of Physics, Lafayette College, Easton, PA 18042, USA}
\author[0000-0001-5465-2889]{Timothy T. Pennucci}
\altaffiliation{NANOGrav Physics Frontiers Center Postdoctoral Fellow}
\affiliation{National Radio Astronomy Observatory, 520 Edgemont Road, Charlottesville, VA 22903, USA}
\affiliation{Institute of Physics, E\"{o}tv\"{o}s Lor\'{a}nd University, P\'{a}zm\'{a}ny P. s. 1/A, 1117 Budapest, Hungary}
\author[0000-0002-8826-1285]{Nihan S. Pol}
\affiliation{Department of Physics and Astronomy, West Virginia University, P.O. Box 6315, Morgantown, WV 26506, USA}
\affiliation{Center for Gravitational Waves and Cosmology, West Virginia University, Chestnut Ridge Research Building, Morgantown, WV 26505, USA}
\affiliation{Department of Physics and Astronomy, Vanderbilt University, 2301 Vanderbilt Place, Nashville, TN 37235, USA}
\author[0000-0001-5799-9714]{Scott M. Ransom}
\affiliation{National Radio Astronomy Observatory, 520 Edgemont Road, Charlottesville, VA 22903, USA}
\author[0000-0002-5297-5278]{Paul S. Ray}
\affiliation{Space Science Division, Naval Research Laboratory, Washington, DC 20375-5352, USA}
\author[0000-0002-7283-1124]{Brent J. Shapiro-Albert}
\affiliation{Department of Physics and Astronomy, West Virginia University, P.O. Box 6315, Morgantown, WV 26506, USA}
\affiliation{Center for Gravitational Waves and Cosmology, West Virginia University, Chestnut Ridge Research Building, Morgantown, WV 26505, USA}
\author[0000-0002-7778-2990]{Xavier Siemens}
\affiliation{Department of Physics, Oregon State University, Corvallis, OR 97331, USA}
\affiliation{Center for Gravitation, Cosmology and Astrophysics, Department of Physics, University of Wisconsin-Milwaukee,\\ P.O. Box 413, Milwaukee, WI 53201, USA}
\author[0000-0003-1407-6607]{Joseph Simon}
\affiliation{Jet Propulsion Laboratory, California Institute of Technology, 4800 Oak Grove Drive, Pasadena, CA 91109, USA}
\affiliation{Department of Astrophysical and Planetary Sciences, University of Colorado, Boulder, CO 80309, USA}
\author[0000-0002-6730-3298]{Ren\'{e}e Spiewak}
\affiliation{Centre for Astrophysics and Supercomputing, Swinburne University of Technology, P.O. Box 218, Hawthorn, Victoria 3122, Australia}
\author[0000-0001-9784-8670]{Ingrid H. Stairs}
\affiliation{Department of Physics and Astronomy, University of British Columbia, 6224 Agricultural Road, Vancouver, BC V6T 1Z1, Canada}
\author[0000-0002-1797-3277]{Daniel R. Stinebring}
\affiliation{Department of Physics and Astronomy, Oberlin College, Oberlin, OH 44074, USA}
\author[0000-0002-7261-594X]{Kevin Stovall}
\affiliation{National Radio Astronomy Observatory, 1003 Lopezville Rd., Socorro, NM 87801, USA}
\author{Jerry P. Sun}
\affiliation{Department of Physics, Oregon State University, Corvallis, OR 97331, USA}
\author[0000-0002-1075-3837]{Joseph K. Swiggum}
\altaffiliation{NANOGrav Physics Frontiers Center Postdoctoral Fellow}
\affiliation{Department of Physics, Lafayette College, Easton, PA 18042, USA}
\author[0000-0003-0264-1453]{Stephen R. Taylor}
\affiliation{Department of Physics and Astronomy, Vanderbilt University, 2301 Vanderbilt Place, Nashville, TN 37235, USA}
\author{Jacob E. Turner}
\affiliation{Department of Physics and Astronomy, West Virginia University, P.O. Box 6315, Morgantown, WV 26506, USA}
\affiliation{Center for Gravitational Waves and Cosmology, West Virginia University, Chestnut Ridge Research Building, Morgantown, WV 26505, USA}
\author[0000-0002-4162-0033]{Michele Vallisneri}
\affiliation{Jet Propulsion Laboratory, California Institute of Technology, 4800 Oak Grove Drive, Pasadena, CA 91109, USA}
\author[0000-0003-4700-9072]{Sarah J. Vigeland}
\affiliation{Center for Gravitation, Cosmology and Astrophysics, Department of Physics, University of Wisconsin-Milwaukee,\\ P.O. Box 413, Milwaukee, WI 53201, USA}
\author[0000-0002-6020-9274]{Caitlin A. Witt}
\affiliation{Department of Physics and Astronomy, West Virginia University, P.O. Box 6315, Morgantown, WV 26506, USA}
\affiliation{Center for Gravitational Waves and Cosmology, West Virginia University, Chestnut Ridge Research Building, Morgantown, WV 26505, USA}

\collaboration{1000}{The NANOGrav Collaboration}

\noaffiliation

\correspondingauthor{Joseph Simon}
\email{joe.simon@nanograv.org}

\begin{abstract}
We search for an isotropic stochastic gravitational-wave background (GWB) in the $12.5$-year pulsar-timing data set collected by the North American Nanohertz Observatory for Gravitational Waves. Our analysis finds strong evidence of a stochastic process, modeled as a power-law, with common amplitude and spectral slope across pulsars. Under our fiducial model, the Bayesian posterior of the amplitude for an $f^{-2/3}$ power-law spectrum, expressed as the characteristic GW strain, has median $1.92 \times 10^{-15}$ and $5\%$--$95\%$ quantiles of $1.37$--$2.67 \times 10^{-15}$ at a reference frequency of $\fyr = 1 ~\mathrm{yr}^{-1}$; the Bayes factor in favor of the common-spectrum process versus independent red-noise processes in each pulsar exceeds $10,000$.
However, we find no statistically significant evidence that this process has quadrupolar spatial correlations, which we would consider necessary to claim a GWB detection consistent with general relativity. We find that the process has neither monopolar nor dipolar correlations, which may arise from, for example, reference clock or solar system ephemeris systematics, respectively. 
The amplitude posterior has significant support above previously reported upper limits; we explain this in terms of the Bayesian priors assumed for intrinsic pulsar red noise. We examine potential implications for the supermassive black hole binary population under the hypothesis that the signal is indeed astrophysical in nature.
\end{abstract}

\section{Introduction}
\label{sec:intro}
% !TEX root = nanograv_12p5yr_gwb.tex
 
Pulsar-timing arrays (PTAs; \citealt{saz78, det79, fb90}) seek to detect very-low-frequency ($\sim 1$--$100$ nHz) gravitational waves (GWs) by monitoring the spatially correlated fluctuations induced by the waves on the times of arrival of radio pulses from millisecond pulsars (MSPs). % in the Galaxy.
The dominant source of gravitational radiation in this band is expected to be the stochastic background generated by a cosmic population of supermassive black hole binaries (SMBHBs; \citealt{shm+04, stc+19}).  
Other more speculative stochastic GW sources in the nanohertz frequency range include cosmic strings \citep{smc07, bos18}, phase transitions \citep{cds10,klm+17}, and a primordial GW background (GWB) produced by quantum fluctuations of the gravitational field in the early universe, amplified by inflation \citep{g75, lms+16}.
%The effect of a GWB on the timing data for each pulsar consists of two terms: (i) the "pulsar term," the effect of the metric perturbation at the location of the pulsar, which is uncorrelated from one pulsar to the next; and (ii) the "Earth term," the effect of the metric perturbation at the location of Earth, which is correlated between pulsars according to the Hellings-Downs function \cite{hd83}. However, the cross-correlation coefficients are rather small ($<=0.5$), thus the uncorrelated auto-correlation terms are expected to dominate the signal recovery.

The North American Nanohertz Observatory for Gravitational Waves (NANOGrav; \citealt{ransom+19}) %was formed in 2007 \citep{m13} and 
has been acquiring pulsar-timing data since 2004. NANOGrav is one of three major PTAs along with the European Pulsar Timing Array (EPTA; \citealt{dcl+16}), and the Parkes Pulsar Timing Array (PPTA; \citealt{krh+20}). 
Additionally, there are growing PTA efforts in India \citep{InPTA} and China \citep{CPTA}, as well as some telescope-centered timing programs \citep{MeerTime, CHIMEPulsar}.
In concert, these collaborations support the International Pulsar Timing Array (IPTA; \citealt{pdd+19}).
Over the last decade, PTAs have produced increasingly sensitive data sets, as seen in the steady march of declining upper limits on the stochastic GWB \citep{vhj+11,dfg+13,src+13,ltm+15,srl+15,vlh+16,abb+16,abb+18b}.
It was widely expected that the first inklings of a GWB would manifest in the stagnation of improvement in upper limits, followed by the emergence of a spatially uncorrelated common-spectrum red process in all pulsars, and culminate in the detection of interpulsar spatial correlations with the quadrupolar signature described by \cite{hd83}. 
In practice, it appears that the early indications of a signal may have been obscured by systematic effects due to incomplete knowledge of the assumed position of the solar system barycenter \citep{vts+20}.

In this article, we present our analysis of NANOGrav's newest ``12.5-year" data set \citep[][hereafter \citetalias{aab+20}]{aab+20}. 
We find a strong preference for a stochastic common-spectrum process, modeled as a power-law, in the timing behaviors of all pulsars in the data set. Building on the statistical-inference framework put in place during our GW study of the $11$-year data set \citep[][hereafter \citetalias{abb+18b}]{abb+18b}, we report Bayes factors from extensive model comparisons. 
We find the log$_{10}$ Bayes factor for a spatially uncorrelated common-spectrum process versus independent red-noise processes in each pulsar to range from 2.7 to 4.5, depending on which solar system ephemeris (SSE) modeling scheme we employ.
We model a spatially uncorrelated common-spectrum process to have the same power spectral density across all pulsars in the data set, but with independent realizations in the specific timing behavior of each pulsar.
The evidence is only slightly higher for a common-spectrum process with quadrupolar correlations, with a log$_{10}$ Bayes factor against a spatially uncorrelated common-spectrum process ranging from 0.37 to 0.64, again depending on SSE modeling.
Correspondingly, the Bayesian--frequentist hybrid optimal-statistic analysis \citep{abc+2009,dfg+13,ccs+2015,vite18}, which measures interpulsar correlated power only, is unable to distinguish between different spatially-correlated processes. 
Thus, lacking definitive evidence of quadrupolar spatial correlations, the analysis of this data set must be considered inconclusive with regard to GW detection. 

With an eye toward searches in future, more informative data sets, we perform a suite of statistical tests on the robustness of our findings.
Focusing first on the stochastic common-spectrum process, we examine the contribution of each pulsar to the overall Bayes factor with a dropout analysis \citep{aab+19, vigeland20}, and find broad support among the pulsars in the data set.
Moving on to spatial correlations, we build null background distributions for the correlation statistics by applying random phase shifts and sky scrambles to our data \citep{cs16,tlb+17}, and find that the no-correlations hypothesis is rejected only mildly, with $p$ values $\sim 5\%$ (i.e., $2\sigma$).

The posterior on the amplitude of the common-spectrum process, \Acp, modeled with an $f^{-2/3}$ power-law spectrum, has a median of $1.9 \times 10^{-15}$, with $5\%$--$95\%$ quantiles of $1.4$--$2.7 \times 10^{-15}$ at a reference frequency of $\fyr = 1 ~\mathrm{yr}^{-1}$,  based on a log-uniform prior and using the latest JPL SSE (DE438, \citealt{de438}), which we take as our fiducial model in this paper.
This refined version of the SSE incorporates data from the NASA orbiter Juno\footnote{\href{https://www.missionjuno.swri.edu}{https://www.missionjuno.swri.edu}}, and claims a Jupiter orbit accuracy a factor of 4 better than previous SSEs, which is promising given that our previous analysis showed that errors in Jupiter's orbit dominated the SSE-induced GWB systematics \citep{vts+20}.

The fact that the median value of \Acp\ is higher than the $95\%$ upper limit reported for the $11$-year data set, $\Agw < 1.45 \times 10^{-15}$ \citepalias{abb+18b}, requires explanation.
While many factors contribute to this discrepancy, simulations show that the standard PTA data model (and most crucially, the uniform priors on the amplitude of pulsar-intrinsic red-noise processes) can often yield Bayesian upper limits lower than the true GWB level, by shifting GWB power to pulsar red noise \citep{hazboun:2020b}. Once all factors are taken into account, the data sets can be reconciled.
However, this accounting suggests that the astrophysical interpretation of past Bayesian upper limits from PTAs may have been overstated. 
Indeed, it is worth noting that, while the source of the common-spectrum process in this data set remains unconfirmed, the posterior on \Acp\ is compatible with many models for the GWB that had previously been deemed in tension with PTA analyses.

This paper is laid out as follows:
Sec.\ \ref{sec:obs} describes the $12.5$-year data set.
Our data model is presented in Sec.\ \ref{sec:data_analysis}.
In Sec.\ \ref{sec:results}, we report on our search for a common-spectrum process in the data set and present the results from our extensive exploration for inter-pulsar correlations. 
Sec.\ \ref{sec:detect} contains a suite of statistical checks on the significance of our detection metrics. 
In Sec.\ \ref{sec:discussion}, we discuss the amplitude of the recovered process, addressing both the discrepancies with previous published upper limits and the potential implications for the SMBHB population, and we conclude with our expectations for future searches.

\pagebreak
\section{The 12.5-year Data Set}
\label{sec:obs}
% !TEX root = nanograv_12p5yr_gwb.tex

The NANOGrav $12.5$-year data set has been released using two separate and independent analyses. The \emph{narrowband} analysis, consisting of the time of arrival (TOA) data and pulsar timing models presented in \citetalias{aab+20}, is very similar in its form and construction to our previous data sets in which many TOAs were calculated within narrow radio-frequency bands for data collected simultaneously across a wide bandwidth. A separate ``wideband'' analysis \citep{12yr_wideband} was also performed in which a single TOA is extracted from broadband observations. Both versions of the data set are publicly available online\footnote{\href{http://data.nanograv.org}{http://data.nanograv.org}}.
The data set consists of observations of 47 MSPs made between July 2004 and June 2017. 
This is the fourth public NANOGrav data set, and adds two MSPs and 1.5 years of observations to the previously released $11$-year data set \citepalias{abb+18a}.
Only pulsars with a timing baseline greater than three years are used in our GW analyses \citep[][hereafter \citetalias{abb+16}]{abb+16}, thus all results in this paper are based on the 45 pulsars that meet that criteria. This is a significant increase from the analyses in \citetalias{abb+18b}, which used 34 pulsars, and the analyses in \citetalias{abb+16}, which used 18. 
Additionally, it is crucial to note that the $12.5$-year data set is more than just an extension of the $11$-year---changes to the data processing pipeline, discussed below, have improved the entire span of the data.
In the following section, we briefly summarize the instruments, observations, and data reduction process for the $12.5$-year data set. 
A more detailed discussion of the data set can be found in \citetalias{aab+20}.

\subsection{Observations}

We used the 305-m Arecibo Observatory (Arecibo or AO) and the 100-m Green Bank Telescope (GBT) to observe the pulsars. 
Arecibo observed all sources that lie within its declination range ($0^\circ < \delta < +39^\circ$), while GBT observed those sources that lie outside of Arecibo's declination range, plus PSRs J1713+0747 and B1937+21. Most sources were observed approximately once per month. 
Six pulsars were observed weekly as part of a high-cadence observing campaign, which began at the GBT in 2013 and at AO in 2015 with the goal of improving our sensitivity to individual GW sources \citep{blf2011,cal+2014}: PSRs J0030+0451, J1640+2224, J1713+0747, J1909$-$3744, J2043+1711, and J2317+1439.

Early observations were recorded using the ASP and GASP systems at Arecibo and GBT, respectively, which sampled bandwidths of 64 MHz \citep{Demorest2007}. Between 2010 and 2012, we transitioned to wideband systems (PUPPI at Arecibo and GUPPI at GBT) which can process up to 800 MHz bandwidths \citep{DuPlain2008, Ford2010}. 
At most observing epochs, the pulsars were observed with two different wide band receivers covering different frequency ranges in order to achieve good sensitivity in the measurement of pulse dispersion due to the interstellar medium (ISM). At Arecibo, the pulsars were observed using the 1.4 GHz receiver plus either the 430 MHz receiver or 2.1 GHz receiver, depending on the pulsar's spectral index and timing characteristics. (Early observations of one pulsar also used the 327 MHz receiver.) At GBT, the monthly observations used the 820 MHz and 1.4 GHz receivers. However these two separate frequency ranges were not observed simultaneously; instead, the observations were separated by a few days. 
The weekly observations at GBT used only the 1.4 GHz receiver.

\subsection{Processing and Time-of-arrival Data}

Most of the procedures used to reduce the data, generate the TOAs, and clean the data set were similar to those used to generate previous NANOGrav data sets (\citetalias{abb+15}, \citetalias{abb+18a}); however, several new steps were added. 
We improved the data reduction pipeline by removing low-amplitude artifact images from the profile data that are caused by small mismatches in the gains and timing of the interleaved analog-to-digital converters in the backends. 
We also excised radio frequency interference (RFI) from the calibration files as well as the data files.

We used the same procedures as in \citetalias{abb+15} and \citetalias{abb+18a} to generate the TOAs from the profile data. As we have done in previous data sets, we cleaned the TOAs by removing RFI, low signal-to-noise TOAs \citepalias{abb+15}, and outliers \citepalias{abb+18a}. Compared to previous data sets, we reorganized and systematized the TOA cleaning and timing-model parameter selection processes to improve consistency of processing across all pulsars. We also performed a new test where observing epochs were removed one by one to determine whether removing a particular epoch significantly changed the timing model. This is essentially an outlier analysis for observing epochs rather than individual TOAs.

\subsection{Timing Models and Noise Analysis}\label{sec:timing_noise}

For each pulsar, the cleaned TOAs were fit to a timing model that described the pulsar's spin period and spin period derivative, sky location, proper motion, and parallax. For binary pulsars, the timing model also included five Keplerian binary parameters, and additional post-Keplerian parameters if they improved the timing fit as determined by an $F$-test. We modeled variations in the pulse dispersion as a piecewise constant through the inclusion of DMX parameters (\citetalias{abb+15}, \citealt{jml+2017}). The timing model fits were primarily performed using the \tempo timing software, and the software packages \tempotwo and \pint were used to check for consistency. The timing model fits were done using the TT(BIPM2017) timescale and the JPL SSE model DE436 \citep{de436}. 
The latest JPL SSE (DE438, \citealt{de438}), which we take as our fiducial model for the analyses in this paper, was not available when TOA processing was being done. However, this does not affect the results presented later, as the corresponding changes in the timing parameters are well within their linear range, which is marginalized away in the analysis \citepalias{abb+15, abb+16}.

We modeled noise in the pulsars' residuals with three white-noise components plus a red noise component. The white noise components are EQUAD, which adds white noise in quadrature; ECORR, which describes white noise that is correlated within the same observing epoch but uncorrelated between different observing epochs; and EFAC, which scales the total template-fitting TOA uncertainty after the inclusion of the previous two white noise terms. For all of these components, we used separate parameters for every combination of pulsar, backend, and receiver.

Many processes can produce red noise in pulsar residuals. The stochastic GWB appears in the residuals as red noise, however it appears specifically correlated between different pulsars \citep{hd83}. Other astrophysical sources of red noise include spin noise, pulse profile changes, and imperfectly modeled dispersion measure variations \citep{cordes2013, lcc+2017, jml+2017}. These red noise sources are unique to a given pulsar. There are also potential terrestrial sources of red noise, including clock errors and ephemeris errors \citep{thk+2016}, which are correlated differently than the GWB. We model the intrinsic red noise of each pulsar as a power-law, similar to the GWB (see Sec.\ \ref{sec:gw_spectrum}). 

The changes to the data processing procedure described above significantly improved the quality of the data. In order to quantify the effect of these changes, we produced an ``$11$-year slice'' data set by truncating the $12.5$-year data set at the MJD corresponding to the last observation in the $11$-year data set, and compared the results of a full noise analysis of this data set to those for the $11$-year data set.
As discussed in \citetalias{aab+20}, we found a reduction in the amount of white noise in the $11$-year slice compared to the $11$-year data set.
However, we also found that the red noise changed for many pulsars. Specifically, there is a slight preference for a steeper spectral index across most of the pulsars, indicating that for some pulsars the reduction in white noise produced an increased sensitivity to low-frequency red noise processes, like the GWB.

\section{Data Model}
\label{sec:data_analysis}
% !TEX root = nanograv_12p5yr_gwb.tex

%We should only describe the new features used in this analysis, and point back to the $11$-year stochastic paper for a broader discussion of enterprise and our PTA analysis framework.

The statistical framework for the characterization of noise processes and GW signals in pulsar-timing data is well documented \citepalias[see e.g.,][]{abb+16,abb+18b}.
In this section we give a concise description of our probabilistic model of the 12.5-year data set, focusing on the differences from earlier studies.
The model attempts to represent every known deterministic and stochastic source of timing residuals that could be interpreted as GWs: it extends the individual timing models of the pulsars (discussed in Sec.\ \ref{sec:timing_noise}) by adding common-spectrum processes with specific correlation structures between pulsars.
In Sec.\ \ref{sec:gw_spectrum} we define our spectral models of time-correlated (red) processes, which include pulsar-intrinsic red noise and the GWB; in Sec.\ \ref{sec:ptalike} we list the combinations of time-correlated processes included in our Bayesian model-comparison trials; in Sec.~\ref{sec:sse} we discuss our prescriptions for the solar system ephemeris.
Our Bayesian and frequentist techniques of choice will be described alongside our results in Secs.\ \ref{sec:results} and \ref{sec:detect}, with more technical details in \autoref{app:methods} and \autoref{app:software}.

\subsection{Models of time-correlated processes}
\label{sec:gw_spectrum}

\begin{figure*}[t]
\begin{center}
    \includegraphics[width=0.9\textwidth]{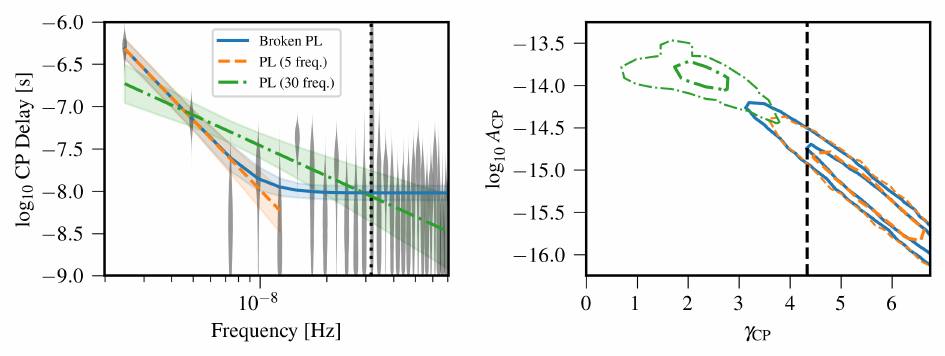}
\end{center}
    \caption{Posteriors for a common-spectrum process in \citetalias{aab+20}, as recovered with four models: free-spectrum (gray violin plots in left panel), broken power law (solid blue lines and contours), 5-frequency power law (dashed orange lines and contours), and 30-frequency power law (dot-dashed green lines and contours). 
    In the left panel, the violin plots show marginalized posteriors of the equivalent amplitude of the sine-cosine Fourier pair (i.e., $\sqrt{S(f)/T}$, with units of seconds) at the frequencies on the horizontal axis; the lines show the mean reconstructed power laws in the left panel, and the 1$\sigma$ (thicker) and 2$\sigma$ posterior contours for the amplitude and spectral slope in the right panel.
    In the left panel, the shaded regions trace $\pm 1\sigma$ ranges for the common-spectrum process power as a function of frequency, as implied by the Bayesian posteriors for the power-law parameters.
   	The dotted vertical line in the left panel sits at $\fyr = 1 \mathrm{yr}^{-1}$, where PTA sensitivity is reduced by the fitting of timing-model parameters; the corresponding free-spectrum amplitude posterior is unconstrained.
   	The dashed vertical line in the right panel sits at $\gamma = 13/3$, the expected value for a GWB produced by a population of inspiraling SMBHBs. 
   	For both the broken power law and 5-frequency power law models, the amplitude (\Acp) posterior shown on the right is extrapolated from the lowest frequencies to the reference frequency $\fyr$.
	We observe that the slope and amplitude of the 30-frequency power law are driven by higher-frequency noise, whereas the 5-frequency power law recovers the low-frequency GWB-like slope of the free spectrum and broken power law. 
	\label{fig:cRN_spectrum}}
\end{figure*}

The principal results of this paper are referred to a fiducial power-law spectrum of the characteristic GW strain:
\begin{equation}
	h_c(f) = A_\mathrm{GWB} \left( \frac{f}{\fyr} \right)^{\!\!\alpha},
	\label{eq:specdef}
\end{equation}
with $\alpha = -2/3$ for a population of inspiraling SMBHBs in circular orbits whose evolution is dominated by GW emission \citep{p01}.
We performed our analysis in terms of the timing-residual cross-power spectral density
\begin{equation}
    S_{ab}(f) = \Gamma_{ab}\,\frac{{A_\mathrm{GWB}}^2}{12\pi^2} \left(\frac{f}{\fyr}\right)^{\!\!-\gamma}\, \fyr^{-3}.
	\label{eq:toaspec}
\end{equation}
where $\gamma = 3 - 2\alpha$ (so the fiducial SMBHB $\alpha=-2/3$ corresponds to $\gamma=13/3$), and where $\Gamma_{ab}$ is the overlap reduction function (ORF), which describes average correlations between pulsars $a$ and $b$ in the array as a function of the angle between them.
For an isotropic GWB, the ORF is given by \citet{hd83} and we refer to it casually as ``quadrupolar'' or ``HD'' correlations.

Other spatially correlated effects present with different ORFs. Systematic errors in the solar system ephemeris have a dipolar ORF, $\Gamma_{ab} = \cos \zeta_{ab}$, where $\zeta_{ab}$ represents the angle between pulsars $a$ and $b$. While errors in the timescale (the ``clock") have a monopolar ORF, $\Gamma_{ab} = 1$. 
Pulsar-intrinsic red noise is also modeled as a power-law, however, in that case there is no ORF. The $A_\mathrm{GWB}$ in Eq.\ \eqref{eq:toaspec} is replaced with an $A_\mathrm{red}$, and $\gamma$ with $\gamma_\mathrm{red}$. There is a separate ($A_\mathrm{red}$, $\gamma_\mathrm{red}$) pair for each pulsar in the array. 

As in \citetalias{abb+16} and \citetalias{abb+18b}, we implemented stationary Gaussian processes with a power-law spectrum in rank-reduced fashion, by approximating them as a sum over a sine--cosine Fourier basis with frequencies $k/T$ and prior (weight) covariance $S_{ab}(k/T) / T$, where $T$ is the span between the minimum and maximum TOA in the array \citep{2014PhRvD..90j4012V}. We use the same basis vectors to model all red noise in the array, both pulsar-intrinsic noise and global signals, like the GWB. Using a common set of vectors helps the sampling, and reduces the likelihood computation time. 
In previous work, the number of basis vectors was chosen to be large enough (with $k = 1, \ldots, 30$) that inference results (specifically the Bayesian upper limit) for a common-spectrum signal became insensitive to adding more components.
However, doing so has the disadvantage of potentially coupling white noise to the highest-frequency components of the red-noise process, thus biasing the recovery of the putative GWB, which is strongest in the lowest-frequency bins.

For this paper, we revisit the issue and set the number of frequency components used to model common-spectrum signals to five, on the basis of theoretical arguments backed by a preliminary analysis of the data set. We begin with the former. By computing a strain spectrum sensitivity curve for the 12.5-year data set using the \textsc{hasasia} tool \citep{hasasia} and obtaining the signal-to-noise ratio (S/N) of a $\gamma=13/3$ power-law GWB, we observed that the five lowest frequency bins contribute $99.98\%$ of the S/N, with the majority coming from the first bin.
We also injected a $\gamma=13/3$ power-law GWB into the 11-year data set \citetalias{abb+18a}, and measured the response of each frequency using a 30-frequency free spectrum model, in which we allowed the variance of each sine--cosine pair in the red-noise Fourier basis to vary independently. We observed that the lowest few frequencies are the first to respond as we raised the GWB amplitude from undetectable to detectable levels (see \autoref{fig:largest_inj_power_ratio}). The details of this injection analysis are described in \autoref{app:inject}.

Moving on to empirical arguments, in \autoref{fig:cRN_spectrum} we plot the power-spectrum estimates for a spatially-uncorrelated common-spectrum process in the 12.5-year dataset, as computed for:
\begin{itemize}[leftmargin=*]
  \item a free-spectrum model (gray violin plots),
  \item variable-$\gamma$ power-law models [Eq.\ \eqref{eq:toaspec} with $A_\mathrm{GWB} = \Acp$ and $\Gamma_{ab} = \delta_{ab}$] with five and 30 frequency components (dashed lines, showing maximum a posteriori values, as well as 1-$\sigma$/2-$\sigma$ posterior contours), and 
  \item a broken power-law model (solid lines), given by \\
\begin{equation}
S(f) = \frac{A_\mathrm{CP}^2}{12 \pi^2} \left(  \frac{f}{f_\mathrm{yr}} \right)^{-\gamma} \left(1+\left( \frac{f}{f_\mathrm{bend}}\right)^{1/\kappa} \right)^{\kappa (\gamma - \delta)} \, \fyr^{-3},
\label{eq:broken}
\end{equation}
where $\gamma$ and $\delta$ are the slopes at frequencies lower and higher than $f_\mathrm{bend}$, respectively, and $\kappa$ controls the smoothness of the transition. In this paper, we set $\delta=0$ to appropriately capture the white noise coupled at higher frequencies and $\kappa=0.1$, which is small enough to contain the transition between slopes to within an individual frequency bin. 
\end{itemize}
Both the free spectrum and the broken power law capture a steep red process at the lowest frequencies, accordant with expectations for a GWB, which is accompanied by a flatter ``forest'' at higher frequencies. The 30-frequency power law is impacted by power at high frequencies (where we do not expect any detectable contributions from a GWB) and adopts a low spectral index that does not capture the full power in the lowest frequencies. By contrast, the five-frequency power law agrees with the free spectrum and broken power law in recovering a steep-spectral process. 

The problem of pulsar-intrinsic excess noise leaking into the common-spectrum process at high frequencies has already been discussed for the 9- and 11-year NANOGrav data sets \citep{aab+19,aab+20bwm,hazboun:2020a}, and we are addressing it through the creation of individually adapted noise models for each pulsar \citep{simon20}.
For this paper, we find a simpler solution in limiting all common-spectrum models to the five lowest frequencies. By contrast, we used 30 frequency components for all rank-reduced power-law models of pulsar-intrinsic red noise\footnote{The Fourier basis is still built on frequencies $k/T$ where $T$ is the maximum time span between TOAs in the array, and the same basis vectors are still used for all red noise models.}, which is consistent with what is used in individual pulsar noise analyses and in the creation of the data set.

\subsection{Models of spatially correlated processes}
\label{sec:ptalike}

% !TEX root = nanograv_12p5yr_gwb.tex

\newcommand{\cc}{$\checkmark$}

\begin{table}[t]
\begin{center}
\label{tab:modeltab}
\caption{Data models.}
\centering
\begin{scriptsize}
\begin{tabular}{l|c|cccc|ccc}
\hline \hline
 \multicolumn{1}{l|}{\textit{NG11gwb labels}} & {\it 1} & {\it 2A} & {\it 2B} & {\it 2D} & {\it 3A} & {\it (new)} & {\it 3B} & {\it 3D} \\
\hline 
 spatial &  & \multicolumn{4}{c|}{single common-} & \multicolumn{3}{c}{two common-} \\
 correlations &  & \multicolumn{4}{c|}{spectrum process} & \multicolumn{3}{c}{spectrum processes} \\
\hline
 $\bullet$ uncorrelated	&	   & \cc &	   &	 &	   & \cc &     &	 \\
 $\bullet$ dipole	 		&	   &	 & \cc &	 &	   &     & \cc &	 \\
 $\bullet$ monopole		&	   &	 &	   & \cc &	   &     &     & \cc \\
 $\bullet$ HD		 		&	   &	 &     &	 & \cc & \cc & \cc & \cc \\
\hline 
 pulsar-intrinsic		& \multirow{2}{*}{\cc} & \multirow{2}{*}{\cc} & \multirow{2}{*}{\cc} & \multirow{2}{*}{\cc} & \multirow{2}{*}{\cc} & \multirow{2}{*}{\cc} & \multirow{2}{*}{\cc} & \multirow{2}{*}{\cc} \\ 
 red-noise	 &	   &	 &     &	 &  &  &  &  \\
\hline \hline
\end{tabular}
\end{scriptsize}
\end{center}
 \tablecomments{The data models analyzed in this paper are organized by the presence of spatially-correlated common-spectrum noise processes. Model names are added for a direct comparison to the naming scheme employed in \citetalias{abb+18b}.}
\end{table}

We analyzed the 12.5-year data set using a hierarchy of data models, which are compared in Bayesian fashion by evaluating the ratios of their evidence.
All models include the same basic block for each pulsar, consisting of measurement noise, timing-model errors, pulsar-intrinsic white noise, and pulsar-intrinsic red noise described by a 30-frequency variable-$\gamma$ power law; but they differ by the presence of one or two red-noise processes that appear in all pulsars with the same spectrum.
As in previous work \citepalias{abb+16, abb+18b}, we fixed all pulsar-intrinsic white noise parameters to their maximum in the posterior probability distribution recovered from single-pulsar noise studies for computational efficiency. 

The models are listed in \autoref{tab:modeltab}, which also reports their labels as used in \citetalias{abb+18b}.
The most basic variant (model 1 in \citetalias{abb+18b}) includes measurement noise and pulsar-intrinsic processes alone. 

The next group of four models includes a single common-spectrum red-noise process. The first among them (model 2A of \citetalias{abb+18b}) features a GWB-like red-noise process with common spectrum, but \emph{without} HD correlations. Because we expect the correlations to be much harder to detect than the diagonal $S_{aa}$ terms in Eq.\ \eqref{eq:toaspec}, due to the values of the HD ORF ($\Gamma_{ab}$) being less than or equal to $0.5$, and because the corresponding likelihood, which does not include any correlations, is very computationally efficient, this model has been the workhorse of PTA searches. However, the positive identification of a GWB will require evidence of a common-spectrum process with HD correlations, which also belongs to this group (model 3A of \citetalias{abb+18b}). The group is rounded out by common-spectrum processes with dipolar and monopolar spatial correlations, which may represent SSE and clock anomalies. For a convincing GWB detection, we expect the data to favor HD correlations strongly over dipolar, monopolar, or no spatial correlations.

The last group includes an additional common-spectrum red-noise process on top of the GWB-like common-spectrum, HD-correlated process. The second process is taken to have either no spatial correlations, dipolar correlations, or monopolar correlations.

\subsection{Solar System Ephemeris} \label{sec:sse}

In the course of the GWB analysis of NANOGrav's $11$-year data set \citepalias{abb+18b}, we determined that GW statistics were surprisingly sensitive to the choice of solar-system ephemeris (SSE), and we developed a statistical treatment of SSE uncertainties (\bayesephem, \citealt{vts+20}), designed to harmonize GW results for SSEs ranging from JPL's DE421 (published in 2009, and based on data up to 2007) to DE436 (published in 2016, and based on data up to 2015).

This was a rather conservative choice: it would be reasonable to expect that more recent SSEs, based on larger data sets and on more sophisticated data reduction, would be more accurate---an expectation backed by the (somewhat fragmentary) error estimates offered by SSE compilers.
However, our analysis showed that errors in Jupiter's orbit (which create an apparent motion of the solar system barycenter and therefore a spurious R{\o}mer delay) dominate the GWB systematics and that Jupiter's orbit has been adjusted across DE421--DE436, by amounts ($\lesssim$ 50 km) comparable to or larger than stated uncertainties.
Thus we decided to err on the side of caution, with the understanding that the Bayesian marginalization over SSE uncertainties would subtract power from the putative GWB process, as confirmed by simulations \citep{vts+20}.

Luckily, these circumstances have since changed. Jupiter's orbit is being refined with data from NASA orbiter Juno: the latest JPL SSE (DE438, \citealt{de438}) fits range and VLBI measurements from six perijoves, and claims orbit accuracy a factor four better than previous SSEs (i.e., $\lesssim$ 10 km).
In addition, the longer time span of the $12.5$-year data set \citepalias{aab+20} reduces the degeneracy between a GWB and Jupiter's orbit \citep{vts+20}.
Accordingly, we adopt DE438 as the fiducial SSE for the results reported in this paper. For completeness and verification, we report also statistics obtained with \bayesephem, adopting the same treatment of \citetalias{abb+18b}; and with the SSE INPOP19a \citep{INPOP19a}, which incorporates range data from nine Juno perijoves.

The DE438 and INPOP19a Jupiter orbit estimates are not entirely compatible, because the underlying data sets do not overlap completely and are weighted differently; nevertheless, the orbits differ in ways that affect GWB results only slightly, which further increases our confidence in DE438.
In our analysis, we used DE438 and INPOP19a without uncertainty corrections: while it is technically straightforward to constrain \bayesephem\ using the orbital-element covariance matrices provided by the SSE authors, the resulting orbital perturbations are so small that GW results are barely affected \citep{vts+20}.

\vspace{1in}
\section{Gravitational Wave Background Estimates}
\label{sec:results}
% !TEX root = nanograv_12p5yr_gwb.tex

Our Bayesian analysis of the 12.5-year data set shows definitive evidence for the presence of a time-correlated stochastic process with a common amplitude $A_\mathrm{CP}$ and a common spectral index $\gamma_\mathrm{CP}$ across all pulsars.
Given this finding, we do not quote an upper limit on a GWB amplitude as in \citetalias{abb+16} and \citetalias{abb+18b}, but rather report the median value and $90\%$ credible interval of $A_\mathrm{CP}$, as well as the log$_{10}$ Bayes factor for a common-spectrum process vs.\ pulsar-intrinsic red noise only.
Further details onfour Bayesian methodology can be found in \autoref{app:methods}.
In addition, we characterize the evidence for HD correlations, which we take as the crucial marker of GWB detection, by obtaining the Bayes factors between the models of \autoref{tab:modeltab}. 

Our results are presented in Sec.\ \ref{sec:model_compare}, and summarized in \autoref{fig:posteriors} and \autoref{fig:bf}.
In Secs.\ \ref{sec:optstat_results} and \ref{sec:ORF_studies} we explore the evidence for spatial correlations further, by way of the optimal statistic \citep{abc+2009,dfg+13,ccs+2015}, and of a novel Bayesian technique that isolates the cross-correlations in the Gaussian-process likelihood.
The statistical significance of our results for both the common-spectrum process and HD correlations is examined in Sec.\ \ref{sec:detect}.

\subsection{Bayesian analysis} \label{sec:model_compare}

\begin{figure}[t]
\begin{center}
    \includegraphics[width=0.9\columnwidth]{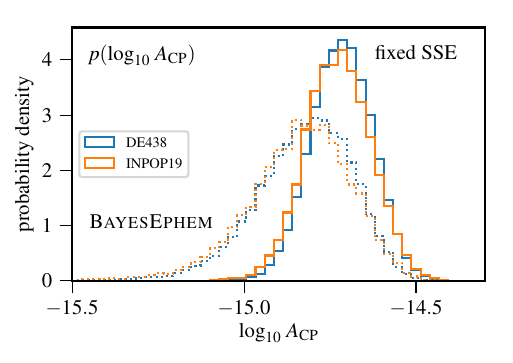}
\end{center}
    \caption{Bayesian posteriors for the ($\fyr = 1 \mathrm{yr}^{-1}$) amplitude \Acp\ of a common-spectrum process, modeled as a $\gamma = 13/3$ power law using only the lowest five component frequencies.
    The posteriors are computed for the NANOGrav $12.5$-year data set using individual ephemerides (solid lines), and \textsc{BayesEphem} (dotted). 
    Unlike similar analyses in \citetalias{abb+18b} and \citet{vts+20}, these posteriors, even those using \bayesephem\, imply a strong preference for a common-spectrum process.
    Results are consistent for both recent SSEs (DE438 and INPOP19a) updated with Jupiter data from the mission Juno.
    SSE corrections remain partially entangled with \Acp. Thus, when \bayesephem\ is applied, the distributions broaden toward lower amplitudes shifting the peak of the distribution by $\sim 20\%$.
    \label{fig:posteriors}}
\end{figure}

\autoref{fig:posteriors} shows marginalized \Acp\ posteriors obtained from the $12.5$-year data using a model that includes pulsar-intrinsic red noise plus a spatially-uncorrelated common-spectrum process with fixed spectral index $\gamma_\mathrm{CP} = 13/3$. Following the discussion of Sec.\ \ref{sec:gw_spectrum}, the common-spectrum process is represented with five sine-cosine pairs. The sine-cosine pairs are modeled to have the same power spectral density, but the values of the coefficients are independent across pulsars. By contrast, in the spatially-correlated models the coefficients are constrained to have the appropriate correlations according to the ORFs.
Under fixed ephemeris DE438, the \Acp\ posterior has median value of $1.92 \times 10^{-15}$ with $5\%$--$95\%$ quantiles at $1.37$--$2.67 \times 10^{-15}$; the INPOP19a posterior is very close---a reassuring finding, given that past versions of the JPL and INPOP SSEs led to discrepant results \citepalias{abb+18b}.

If we allow for \bayesephem\ corrections to DE438, the \Acp\ posterior shifts lower, with median value of $1.53 \times 10^{-15}$ and $5\%$--$95\%$ quantiles at $0.79$--$2.38 \times 10^{-15}$; the posterior for INPOP19a with \bayesephem\ corrections is again very close. It is well understood that \bayesephem\ will absorb power from a common-spectrum process \citep{r19, vts+20}, but we note that this coupling weakens with increasing data set time span: it is weaker here than in the $11$-year analysis, and would be even weaker with $15$ years of data \citep{vts+20}.

\begin{figure*}[t]
\begin{center}
    \includegraphics[width=\textwidth]{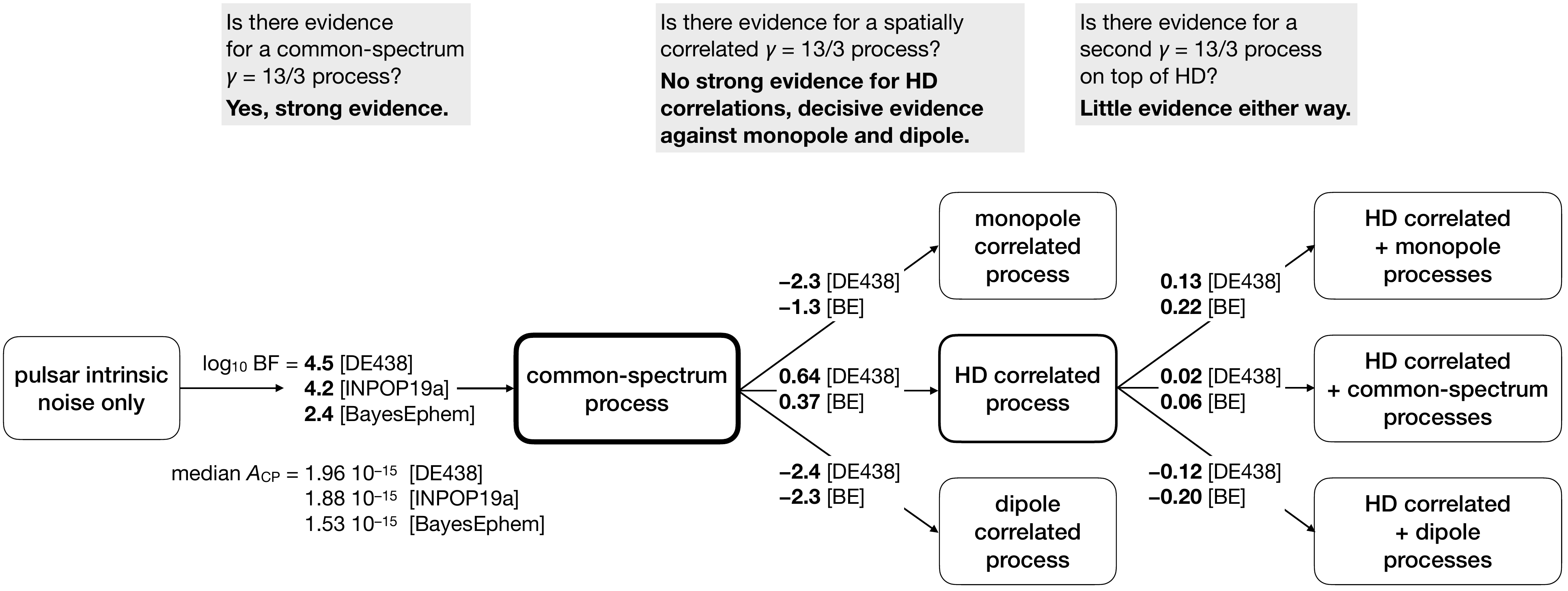}
\end{center}
    \caption{A visual representation of Bayesian model comparisons on the 12.5-yr data set.
    Each box represents a model from \autoref{tab:modeltab}; arrows are annotated with the log$_{10}$ Bayes factor between the two models that they connect, computed for both fixed and \bayesephem-corrected SSE.
    Moving from the left to the right, we find strong evidence for a common-spectrum process, weak evidence for its HD correlations, moderately negative evidence for monopolar or dipolar correlations, and approximately even odds for a second common-spectrum process.
    The log$_{10}$ Bayes factor between any two models can be approximated by summing the values along a path that connects them.
    \label{fig:bf}}
\end{figure*}

% !TEX root = nanograv_12p5yr_gwb.tex

\begin{table*}
 \begin{center}
 \caption{Bayesian model-comparison scores}
	\begin{tabular}{l|c|ccc|ccc}
		\hline\hline
		& \multicolumn{1}{c|}{uncorr. process} & \multicolumn{1}{c}{dipole} 
		& \multicolumn{1}{c}{mono.} & \multicolumn{1}{c|}{HD} 
		& \multicolumn{1}{c}{HD+dip.} & \multicolumn{1}{c}{HD+mono.} 
		& \multicolumn{1}{c}{HD+uncorr.} \\
ephemeris	 	& \multicolumn{1}{c|}{vs.\ noise-only} 
		& \multicolumn{3}{c|}{vs.\ uncorrelated process}
		& \multicolumn{3}{c}{vs.\ HD correlated process} \\
		\hline
DE438           			& 4.5(9) & $-$2.4(2) & $-$2.3(2) & 0.64(1) & $-$0.116(4)  & 0.126(4) & 0.0164(1) \\
		\hline
\textsc{BayesEphem}& 2.4(2) & $-$2.3(2) & $-$1.3(1) & 0.371(5) & $-$0.199(5)  & 0.217(6) & 0.0621(4)  \\
		\hline
   \end{tabular}
 \end{center}
 \label{tab:spatial_bfs_single}
  \tablecomments{The log$_{10}$ Bayes factors between pairs of models from \autoref{tab:modeltab} are also visualized in \autoref{fig:bf}.
 All common-spectrum power-law processes are modeled with fixed spectral index $\gamma=13/3$ and with the lowest five frequency components. The digit in the parentheses gives the uncertainty on the last quoted digit.}
\end{table*}

These peaked, compact \Acp\ posteriors are accompanied by large Bayes factors in favor of a spatially uncorrelated common-spectrum process vs.\ pulsar-intrinsic pulsar red noise alone: $\mathrm{log}_{10}$ Bayes factor $= 4.5$ for DE438, and $2.7$ with \bayesephem. 
Next, we assess the evidence for spatial correlations by computing Bayes factors between the models in \autoref{tab:modeltab}. Our results are summarized in \autoref{tab:spatial_bfs_single} and more visually in \autoref{fig:bf}. 
There is little evidence for the addition of HD correlations ($\mathrm{log}_{10}$ Bayes factor $= 0.64$ with DE438, $0.37$ with \bayesephem), and the HD-correlated \Acp\ posteriors are very similar to those of \autoref{fig:posteriors}. 
By contrast, monopolar and dipolar correlations are moderately disfavored ($\mathrm{log}_{10}$ Bayes factor $= -2.3$ and $-2.4$, respectively, with DE438). The monopole is disfavored less under \bayesephem, which may be explained by the \bayesephem-reduced amplitude of the processes.

The evidence for a second common-spectrum process on top of an HD-correlated process is inconclusive. Furthermore, the amplitude posteriors for additional monopolar and dipolar processes display no clear peaks, while the posterior for an additional spatially uncorrelated process shows that power is drawn away from the HD-correlated process (which is understandable given the scant evidence for HD correlations).

We completed the same analyses with a common-spectrum model where $\gamma_\mathrm{CP}$ was allowed to vary. As seen in \autoref{fig:cRN_spectrum}, the posteriors on $\gamma_\mathrm{CP}$, while consistent with $13/3 ~(\approx 4.33)$, are very broad. Under fixed ephemeris DE438, the $\gamma_\mathrm{CP}$ posterior from a spatially uncorrelated process has a median value of $5.52$ with $5\%$--$95\%$ quantiles at $3.76$--$6.78$. The amplitude posterior is larger in this case, but that is due to the inherent degeneracy between \Acp\ and $\gamma$. The evidence for spatial correlations in a varied-$\gamma_\mathrm{CP}$ model is almost identical to that reported in \autoref{tab:spatial_bfs_single}.

Altogether, the smaller Bayes factors in the discrimination of spatial correlations are fully expected, given that spatial correlations are encoded by the cross-terms in the interpulsar covariance matrix, which are subdominant with respect to the self-terms that drive the detection of a common-spectrum process.
Nevertheless, if a GWB is truly present the Bayes factors will continue to increase as data sets grow in timespan and number of pulsars. Indeed, the trends on display here are broadly similar to the results of \citetalias{abb+18b}, but they have become more marked.

\subsection{Optimal statistic} \label{sec:optstat_results}

The optimal statistic \citep{abc+2009,dfg+13,ccs+2015} is a frequentist estimator of the amplitude of an HD-correlated process, built as sum of correlations among pulsar pairs, weighted by the assumed pulsar-intrinsic and interpulsar noise covariances.
It is a useful complement to Bayesian techniques, specifically for the characterization of spatial correlations.
The statistic $\hat{A}^2$ is defined by Eq.\ 7 of \citetalias{abb+18b},
and it is related to the GWB amplitude by $\langle \hat{A}^2 \rangle = \Agw^2$, where the mean is taken over an ensemble of GWB realizations of the same \Agw.
The statistical significance of an observed $\hat{A}^2$ value is quantified by the corresponding signal-to-noise ratio (S/N, see Eq.\ 8 of \citetalias{abb+18b}).

%\begin{equation}
%	\hat{A}^2 = \frac{\sum_{ab} \res_a^T {{\bf P}_a^{-1} \tilde{{\bf S}}_{ab} {\bf P}_b^{-1} \res_b}}{\sum_{ab} \Tr \left( {\bf P}_a^{-1} \tilde{{\bf S}}_{ab} {\bf P}_b^{-1} \tilde{{\bf S}}_{ba} \right) } \,,
%\end{equation}
%where $a,b$ index pulsars in the PTA, 
%$\res_a$ is the vector of timing residuals for pulsar $a$, 
%${\bf P}_a = \left\langle \res_a \res_a^T \right\rangle$ is the autocovariance matrix of the residuals, 
%and $\hat{A}^2 \tilde{{\bf S}}_{ab} = {\bf S}_{ab} = \left\langle \res_a \res_b^T \right\rangle$ 
%is the cross-correlation matrix of the residuals of pulsar $a$ and $b$ ($a \neq b)$. 
%The optimal statistic $\hat{A}^2$ . 
%The average signal-to-noise ratio (SNR) of the optimal statistic is
%\begin{equation}
%	\rho = \hat{A}^2 \left[\sum_{ab} \Tr \left( {\bf P}_a^{-1} \tilde{{\bf S}}_{ab} {\bf P}_b^{-1} \tilde{{\bf S}}_{ba} \right) \right]^{1/2} \,.
%\end{equation}

\begin{figure}[t]
\begin{center}
    \includegraphics[width=0.9\columnwidth]{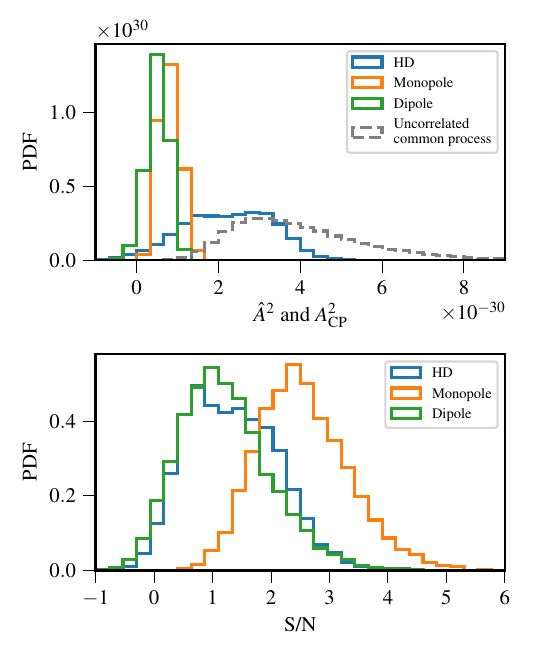}
\end{center}
    \caption{Distributions of the optimal statistic and S/N for HD (blue), monopole (orange), and dipole (green) spatial correlations, as induced by the posterior probability distributions of pulsar-intrinsic red noise parameters in a Bayesian inference run that includes a spatially uncorrelated common-spectrum process. The means of each distribution are the noise-marginalized $\hat{A}^2$ given in \autoref{tab:optstat}. 
    The top panel also shows the posterior of an uncorrelated common red process $\Acp^2$ (dashed gray) from \autoref{fig:posteriors} for comparison. 
    All three cross-correlation patterns are identified in the data with modest significance; but it is only for an HD-correlated process that the amplitude estimate is compatible with the posteriors of \autoref{fig:posteriors}. \label{fig:optstat_spatial_compare}}
\end{figure}

\autoref{tab:optstat} and \autoref{fig:optstat_spatial_compare} summarize the optimal-statistic analysis of the 12.5-year dataset.
As in \citetalias{abb+18b}, we computed two variants of the statistic: a \emph{fixed-noise} version obtained by fixing the pulsar red-noise parameters to their maximum a posteriori values in Bayesian runs that include a spatially uncorrelated common-spectrum process;
and a \emph{noise-marginalized} version \citep{vite18}, which has proved more accurate when pulsars have intrinsic red noise, and which is sampled over 10,000 red-noise parameter vectors drawn from those same posteriors.
For each variant, we computed versions of the statistic tailored to HD, monopolar, and dipolar spatial corrections.
%
% !TEX root = nanograv_12p5yr_gwb.tex

\begin{table}
  \begin{center}
  \caption{Optimal statistic $\hat{A}^2$ and corresponding S/N}
  \begin{tabular}{l|cccc}
  \hline \hline
  & \multicolumn{2}{c}{fixed noise} & \multicolumn{2}{c}{noise marginalized} \\
  correlation & $\hat{A}^2$ & S/N & mean $\hat{A}^2$ & mean S/N \\
  \hline
HD 				& $4\times10^{-30}$ 	& $2.8$ & $2(1)\times10^{-30}$	& $1.3(8)$ \\
monopole 		& $9 \times 10^{-31}$	& $3.4$ & $8(3)\times10^{-31}$	& $2.6(8)$ \\
dipole 			& $9\times10^{-31}$ 	& $2.4$ & $5(3)\times10^{-31}$	& $1.2(8)$ \\
  \hline
  \end{tabular}
  \end{center}
  \label{tab:optstat}
  \tablecomments{The optimal statistic, $\hat{A}^2$, and corresponding S/N are computed from the 12.5-year data set for a HD, monopolar, and dipolar correlated common-process modeled as a power-law with fixed spectral index, $\gamma = 13/3$, using the five lowest frequency components.
  We show fixed intrinsic red-noise and noise-marginalized values. All are computed with fixed ephemeris DE438.}
\end{table}

We recovered similarly low S/N for all three correlation patterns, indicating that the optimal statistic cannot distinguish among them. %This is expected from the coupling that between these correlation patterns \citep{r19}. 
Nevertheless, these results are markedly different from those of \citetalias{abb+18b}, which found no trace of correlations.
The highest S/N is found for the monopolar process, which may seem in conflict with the Bayes factors of \autoref{tab:spatial_bfs_single}; however,\autoref{fig:optstat_spatial_compare} shows that the corresponding amplitude estimate $\hat{A}^2$ is more than a factor of two lower than implied by the \Acp\ posterior, shown there by the dashed curve.
A compatible amplitude estimate is found only for the HD process.
In other words, the optimal-statistic analysis is consistent with the Bayesian analysis. They agree on the presence of an HD-correlated process at the common amplitude indicated by the Bayesian analysis, and both find it strongly unlikely that there are monopolar or dipolar processes of equal amplitude. 
These optimal-statistic results are robust with respect to changing $\gamma$ within the range recovered in \autoref{fig:cRN_spectrum}.

\begin{figure}[t]
\begin{center}
    \includegraphics[width=0.9\columnwidth]{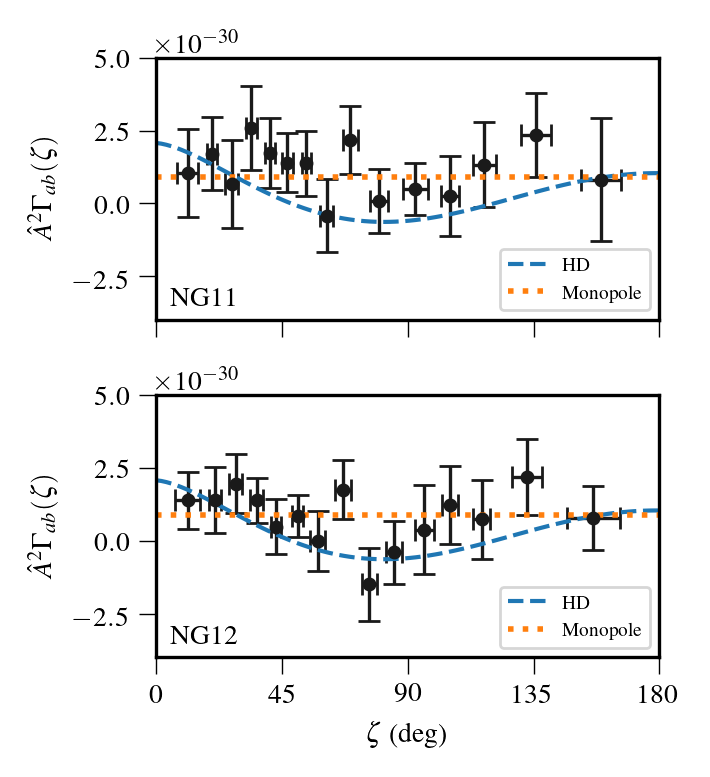}
\end{center}
    \caption{Average angular distribution of cross-correlated power, as estimated with the optimal statistic on the 11-year data set (\textit{top}) and 12.5-year data set (\textit{bottom}). 
    The number of pulsar pairs in each binned point is held constant for each data set. Due to the increase in pulsars in the $12.5$-yr data set, the number of pairs per bin increases accordingly.
    Pulsar-intrinsic red-noise amplitudes are set to their maximum posterior values from the Bayesian analysis, while the SSE is fixed to DE438. 
    The dashed blue and dotted orange lines show the cross-correlated power predicted for HD and monopolar correlations with amplitudes $\hat{A}^2 = 4\times10^{-30}$ and $9 \times 10^{-31}$, respectively. 
    \label{fig:optstat_ang}}
\end{figure}

\autoref{fig:optstat_ang} shows the angular distribution of cross-correlated power for both \citetalias{abb+18a} and \citetalias{aab+20}, as obtained by grouping pulsar pairs into angular-separation bins (with each bin hosting a similar number of pairs). 
The error bars show the standard deviations of angular separations and cross-correlated power within each bin.
The dashed and dotted lines show the values expected theoretically from HD- and monopolar-correlated processes with amplitudes set from the measured $\hat{A}^2$ (the first column of \autoref{tab:optstat}).
While errors are smaller for \citetalias{aab+20} than for \citetalias{abb+18a}, neither correlation pattern is visually apparent.

\subsection{Bayesian measures of spatial correlation} %ORF Shape Studies
\label{sec:ORF_studies}

Inspired by the optimal statistic, we have developed two novel Bayesian schemes to assess spatial correlations. We report here on their application to the 12.5-year data.

\begin{figure}[t]
\begin{center}
    \includegraphics[width=0.9\columnwidth]{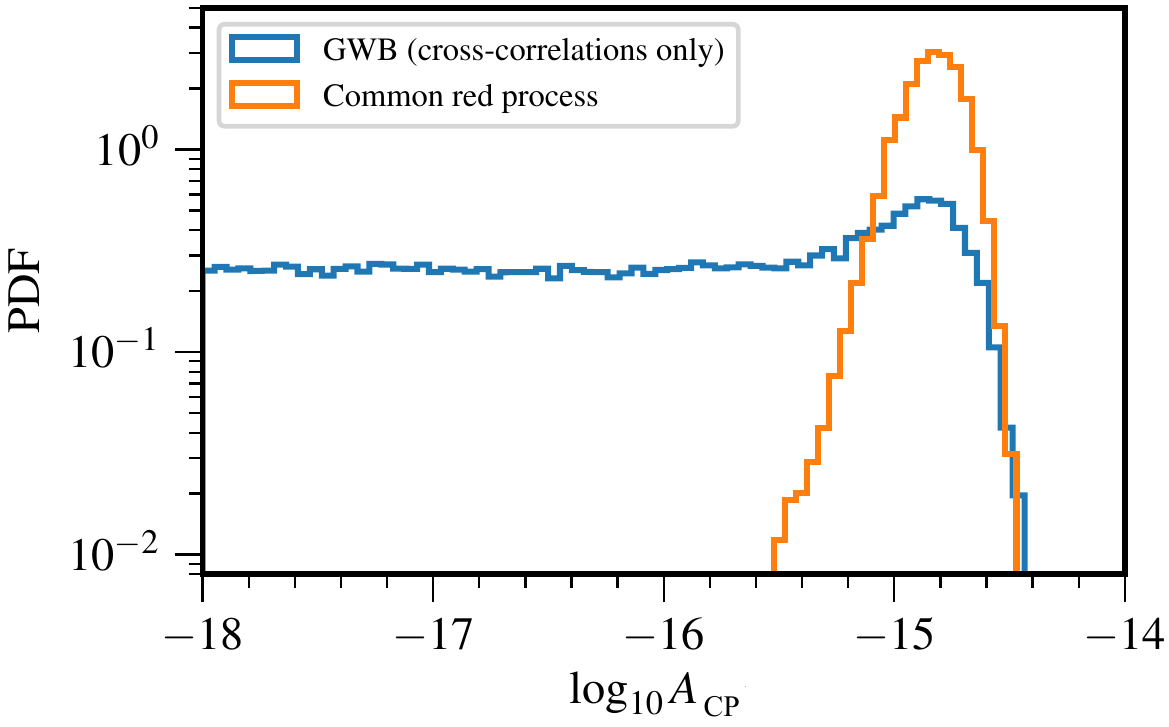}
\end{center}
    \caption{Bayesian amplitude posteriors in a model that includes a common-spectrum process, and an off-diagonal HD-correlated process where all auto-correlation terms are set to zero (see main text of Sec.\ \ref{sec:ORF_studies}). The posteriors shown here are marginalized with respect to each other. The inference run includes \bayesephem.
    \label{fig:crn+offdiag}}
\end{figure}

First, we performed Bayesian inference on a model where the uncorrelated common-spectrum process is augmented with a second HD-correlated process with auto-correlation coefficients set to zero.
In other words, we decouple the amplitudes of the auto- and cross-correlation terms.
The uncorrelated common-spectrum process regularizes the overall covariance matrix, which would not otherwise be positive definite with this new ``off diagonal only'' GWB.
\autoref{fig:crn+offdiag} shows marginalized amplitude posteriors for the diagonal and off-diagonal processes, which appear consistent.
It is however evident that cross correlations carry much weaker information: as a matter of fact, the log$_{10}$ Bayes factor in favor of the additional process (computed \emph{\`a la} Savage--Dickey, see \citealt{d71}) is $0.10 \pm 0.01$ with fixed DE$438$ and $-0.03 \pm 0.01$ under \textsc{BayesEphem}.
These factors are smaller than the HD-vs.-uncorrelated values of \autoref{tab:spatial_bfs_single}, arguably because the off-diagonal portion of the model is given the additional burden of selecting the appropriate amplitude.

\begin{figure}[t]
\begin{center}
    \includegraphics[width=0.9\columnwidth]{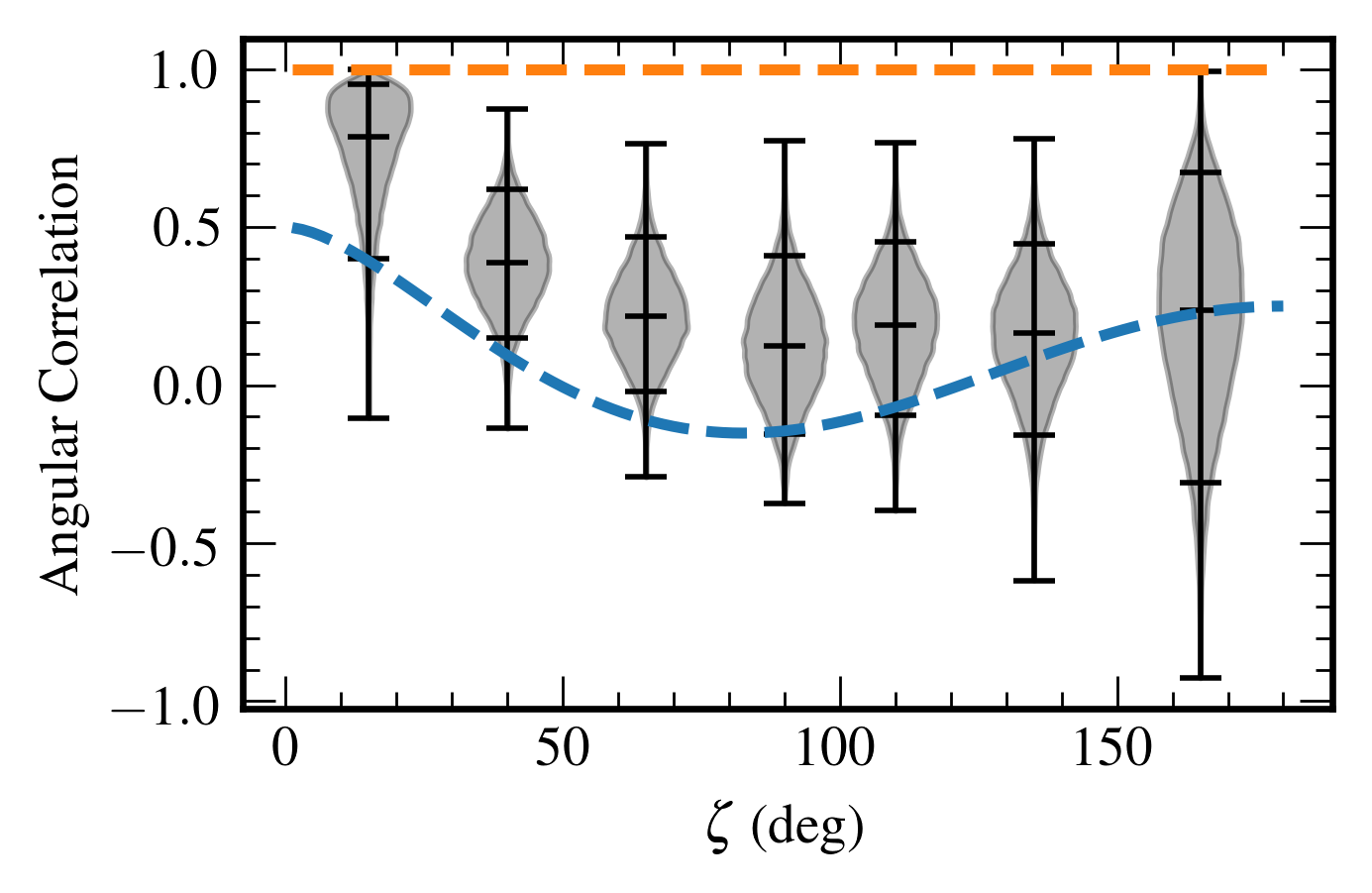}
\end{center}
    \caption{Bayesian reconstruction of interpulsar spatial correlations, parameterized as a seven-node spline. Violin plots show marginalized posteriors for node correlations, with medians, 5\% and 95\% percentiles, and extreme values. The dashed blue line shows the HD ORF expected for a GWB, while the dashed horizontal orange line shows the expected interpulsar correlation signature for a monopole systematic error, (e.g. drifts in clock standards).
    \label{fig:binned_orf}}
\end{figure}

Second, we performed Bayesian inference on a common-spectrum model that includes a parametrized ORF: specifically, inter-pulsar correlations are obtained by the spline interpolation of seven nodes spread across angular separations; node values are estimated as independent parameters with uniform priors in $[-1,1]$ \citep{2013PhRvD..87d4035T}.
\autoref{fig:binned_orf} shows the marginalized posteriors of the angular correlations, and bears direct comparison with \autoref{fig:optstat_ang}. The posteriors, although not very informative, are consistent with the HD ORF, which is over-plotted in the figure. However, they are inconsistent with the monopolar ORF, also overplotted in the figure. This behavior is similar to the evidence reported in \autoref{tab:spatial_bfs_single}. 

\pagebreak
\section{Statistical Significance}
\label{sec:detect}
% !TEX root = nanograv_12p5yr_gwb.tex

As described above, the $12.5$-year data set offers strong evidence for a spatially uncorrelated common-spectrum process across pulsars in the data set, but it favors only slightly the interpretation of this process as a GWB by way of HD inter-pulsar correlations.
In this section we test the robustness of the first statement, by examining the contribution of each pulsar to the overall Bayes factor; and we characterize the statistical significance of the second, by building virtual null distributions for the HD detection statistics.
We expect that studies of both kinds will be important to establishing confidence in future detection claims.

\subsection{Characterizing the evidence for a common-spectrum process across the PTA}
\label{subsec:factorized}

Under a model that includes a noise-like process of common spectrum across all pulsars \emph{without} inter-pulsar correlations, and in the absence of other physical effects linking observations across pulsars (such as ephemeris corrections), the PTA likelihood factorizes into individual pulsar terms:
\begin{equation} \label{eq:faclike}
    p(\{d_j\}_N | \{\vec{\theta}_j\}_N, \Acp) = \prod^N_{j=1} p(d_j|\vec{\theta}_j,\Acp),
\end{equation}
where $d_j$ and $\vec{\theta}_j$ denote the data set and the intrinsic noise parameters for each pulsar $j$, and where \Acp\ denotes the amplitude of the common-spectrum process.

Equation \eqref{eq:faclike} suggests a trivially parallel approach to estimating the \Acp\ posterior: we performed independent inference runs for each pulsar, sampling timing-model parameters, pulsar-intrinsic white-noise parameters, pulsar-intrinsic red-noise parameters, as well as \Acp. We adopted DE438 (without corrections) as the solar-system ephemeris, and we set log-uniform priors for all red-process amplitudes (as seen in \autoref{tab:priors}).
We then obtained $p(\Acp|\{d_j\}_N)$ by multiplying the individual $p(\Acp|d_j)$ posteriors (as represented, e.g., by kernel density estimators), while correcting for the duplication of the prior $p(\Acp)$.

\begin{figure}[t]
\begin{center}
    \includegraphics[width=0.9\columnwidth]{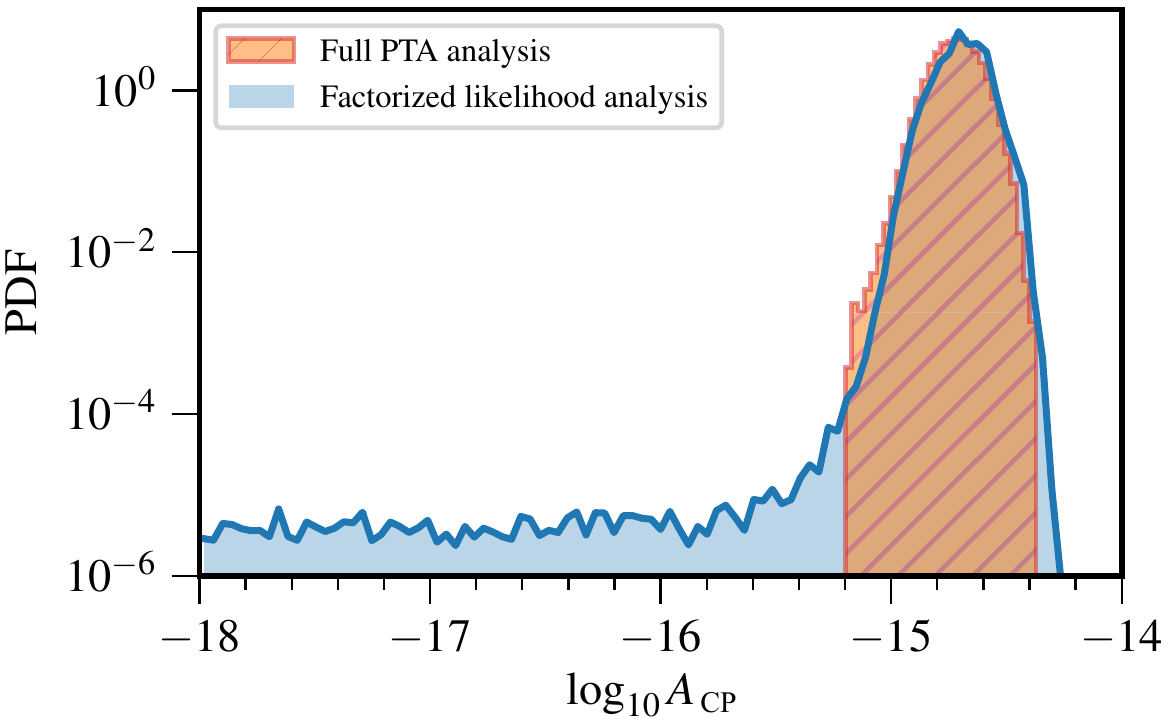}
\end{center}
    \caption{Marginalized \Acp\ posterior of a common-spectrum process modeled with a fixed $\gamma = 13/3$ power law with five component frequencies and no inter-pulsar correlations, as evaluated with full-PTA sampling and with the factorized likelihood approach of Sec.\ \ref{subsec:factorized}. We fixed the ephemeris to DE438 (without corrections), and varied white-noise hyper-parameters for the factorized likelihood, but not in the full-PTA run. Note the logarithmic vertical scale, which emphasizes the very-low-density tail of the distribution; full-PTA sampling has trouble accessing that region because low \Acp\ requires the fine tuning of relatively high $A_\mathrm{red}$ in most pulsars.
  \label{fig:fac_like}}
\end{figure}

As shown in \autoref{fig:fac_like}, the resulting posterior matches the analysis of Sec.\ \ref{sec:results}, while sampling very low \Acp\ values more accurately.
We can then evaluate the $p_\mathrm{all}(\mathrm{CP}) / p_\mathrm{all}(\mathrm{no\,CP})$ Bayes factor in the Savage--Dickey approximation (see \citealt{d71}), obtaining a value $\sim 65,000$, or $\mathrm{log}_{10}$ Bayes factor $\sim 4.8$, which is broadly consistent with the transdimensional sampling estimates reported in \autoref{tab:spatial_bfs_single}.
The agreement of the two distributions in \autoref{fig:fac_like} validates the approximation of fixing pulsar-intrinsic white-noise hyperparameters in the full-PTA analysis, which we accepted for the sake of sampling efficiency.

In a \emph{dropout} analysis \citep{aab+19, vigeland20}, we perform inference on the joint PTA data set, but introduce a binary indicator parameter for each pulsar that can turn off the common-spectrum process term in the likelihood of its data. These indicators are sampled in Monte Carlo fashion with all other parameters. The dropout factor (the number of ``on'' samples divided by ``off'' samples for a pulsar) quantifies the support offered by each pulsar to the common-signal hypothesis. 

\begin{figure*}[t]
  \begin{center}
    \includegraphics[width=0.9\textwidth]{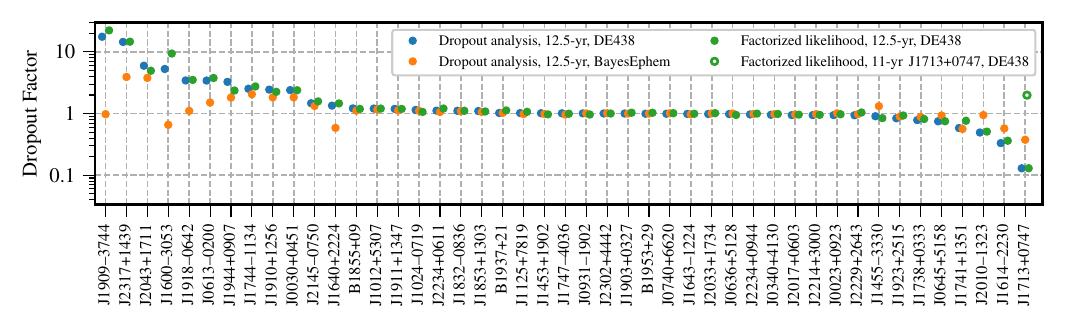}
  \end{center}
    \caption{Characterizing the evidence from each pulsar in favor of a common-spectrum, no-correlations, stochastic process modeled as a $\gamma=13/3$ power law.
    Direct dropout factors (see Eq.\ \eqref{eq:dropout}) from fixed pulsar-intrinsic white-noise, fixed DE438 runs are shown as blue points; they match the estimates from variable white-noise, fixed DE438 factorized likelihoods indicated by green points. The orange points show dropout factors when we include \bayesephem\ corrections. Most of the evidence arises from the ten pulsars on the left, while PSRs J2010$-$1323, J1614$-$2230, and J1713+0747 remain skeptical. All of these effects are diminished by \bayesephem, except for PSR J1713+0747. However, a factorized likelihood analysis using the 11-yr version of PSR J1713+0747 shows modest evidence for the common process, as indicated by the hollow green point. This suggests that an unmodeled noise process in the $12.5$-year version of PSR J1713+0747 is preventing the pulsar from showing evidence for the common-spectrum process.
    \label{fig:dropout_combined}}
\end{figure*}

In this paper, we allow only a single pulsar to drop out at any time in the exploration of the posterior. We performed such dropout runs with fixed pulsar-intrinsic white-noise parameters and fixed ephemeris DE438; the resulting dropout factors are displayed by the blue dots of \autoref{fig:dropout_combined}, sorted by decreasing value. Of the 45 pulsars used in this analysis, roughly 10 have values significantly larger than 1, and (by implication) contribute most of the evidence toward the recovered common-spectrum process, 3 (notably PSR J1713+0747) disfavor that hypothesis, and prefer to ``drop out", while the rest remain agnostic.

The dropout factor for each pulsar $k$ is linked to the \emph{posterior predictive likelihood} for the single-pulsar data set $d_k$, integrated over the  \Acp\ posterior from all other pulsars \citep{wtv19}: 
\begin{multline}
\mathrm{ppl}_k(\mathrm{CP}) =
\int \Bigl[ p(d_k | \vec{\theta}_k,  \Acp) \times \\ p( \Acp | \{d_{j \neq k}\}) \times p(\vec{\theta}_k) \, \mathrm{d}  \Acp \Bigr] \mathrm{d} \vec{\theta}_k.
\end{multline}
If the likelihood factorizes per Eq.\ \eqref{eq:faclike}, then the dropout factor is
\begin{equation}
\label{eq:dropout}
\mathrm{dropout}_k = \frac{p_\mathrm{all}(\mathrm{CP})}{p_k(\mathrm{no\,CP}) p_{j \neq k}(\mathrm{CP})} = \frac{\mathrm{ppl}_k(\mathrm{CP})}{p_k(\mathrm{no\,CP})} \\
\end{equation}
where $p_\mathrm{all}(\mathrm{CP})$ and $p_{j \neq k}(\mathrm{CP})$ denote the Bayesian evidence for the common-spectrum model from all pulsars together, and from all pulsars excluding $k$, respectively; and where $p_k(\mathrm{no\,CP})$ is the evidence for the intrinsic-noise-only model in the data from pulsar $k$.

The posterior predictive likelihood quantifies model support by Bayesian cross validation: namely, the \Acp\ posterior obtained from $n - 1$ pulsars is used to compute the likelihood of the data measured for the excluded pulsar, which acts as an out-of-sample testing data set \citep{wtv19}.
In other words, single-pulsar data sets with dropout factor larger than one can be predicted successfully from the \Acp\ posterior from all other pulsars, lending credence to the common spectrum process model as a whole.
Small dropout factors indicate problematic single-pulsar data sets, or deficiencies in the global model.

Equation \eqref{eq:dropout} can be recast as
\begin{multline}
\mathrm{dropout}_k = \frac{p_k(\mathrm{CP})}{p_k(\mathrm{no\,CP})} \times \\ \int \frac{p(\Acp|\{d_{j \neq k}\}) \, p(\Acp | d_k)}{p(\Acp)} \, \mathrm{d}\Acp,
\end{multline}
which allows the numerical evaluation of dropout factors from factorized likelihoods, where the Bayes factor can be computed \emph{\`a la} Savage--Dickey from the single-pulsar analysis of each pulsar. The resulting $\mathrm{dropout}_k$ estimates are shown as the green dots in \autoref{fig:dropout_combined}, and they agree closely with the direct dropout estimates. 

Unlike the factorized-likelihood approximation, the dropout analysis remains possible when model parameters are included that correlate the likelihoods, such as \bayesephem\ correction coefficients.
Dropout factors for that case are shown as orange dots in \autoref{fig:dropout_combined}, and they can still be interpreted as indicators of the positive or negative evidence contributed by each pulsar toward the common-spectrum process hypothesis. Introducing \bayesephem\ yields reduced factors for the first ten pulsars, consistent with the partial absorption of GW-like residuals into ephemeris corrections \citep{vts+20}. Two of the contrarian pulsars also revert to neutral factors, but PSR J1713+0747 does not.

Altogether, the dropout analysis suggests that the strong evidence for a common-spectrum process originates from more than just a few outliers of NANOGrav pulsars. In \autoref{tab:dropout} we summarize the timing properties of the ten pulsars with dropout factors greater than two. 
As expected, most of the evidence for the common-spectrum process comes from pulsars with longer observing baselines. 
We also note that of the 13 pulsars that have been observed for more than 12 years, 
six have dropout factors greater than two, and only one has a dropout factor significantly less than one (PSR J1713+0747). 
Data sets for three pulsars remain somewhat inconsistent with the consensus. If this trend continues as more data are collected, it will be necessary to explain their behavior either as an expected statistical fluctuation, or as the result of pulsar-specific modeling or measurement issues. Work is ongoing to develop advanced noise models specific to each pulsar \citep{simon20}, which will provide a first quantitative assessment.

In the case of PSR J1713+0747, an unmodeled noise process may indeed be to blame. A factorized likelihood analysis using the version of PSR J1713+0747 in the NANOGrav 11-year data set \citepalias{abb+18a} does show weak evidence for the common process, with a dropout factor of 2.0, indicated by a hollow green circle in \autoref{fig:dropout_combined}. This suggests that some issue with the timing or noise model used to describe the $12.5$-year version of PSR J1713+0747 is causing its anomalously low dropout factor.
This is likely due in some part to the ``second'' chromatic timing event \citep{leg+2018}. An extensive study of PSR J1713+0747's noise property's response to the ``first'' chromatic timing event showed that it took a few years of additional data for the red noise properties of the pulsar to return to ``normal'' \citep{hazboun:2020a}. If this is the primary cause of PSR J1713+0747's behavior in the $12.5$-year data set, then future data sets should show a return to previously measured intrinsic red noise values. In which case, the pulsar would then contribute to any future detection claims.

% !TEX root = nanograv_12p5yr_gwb.tex

\begin{table}
  \begin{center}
  \caption{Timing properties of pulsars with high Dropout Factors.}
  \begin{tabular}{cccc}
  \hline \hline
  Pulsar & Dropout Factor & Obs Time & Timing RMS\footnote{Weighted root-mean-square of epoch-averaged post-fit timing residuals, excluding red noise contributions. See Table 3 of \citetalias{aab+20}.} \\
   & (DE438) & [yrs] & [$\mu$s] \\
  \hline
J1909$-$3744 & 17.6 & 12.7 & 0.061 \\
J2317+1439 & 14.5 & 12.5 & 0.252 \\
J2043+1711 & \hphantom{0}6.0 & \hphantom{0}6.0 & 0.151 \\
J1600$-$3053 & \hphantom{0}5.3 & \hphantom{0}9.6 & 0.245 \\
J1918$-$0612 & \hphantom{0}3.4 & 12.7 & 0.299 \\
J0613$-$0200 & \hphantom{0}3.4 & 12.3 & 0.178 \\
J1944+0907 & \hphantom{0}3.3 & \hphantom{0}9.3 & 0.365 \\
J1744+1134 & \hphantom{0}2.5 & 12.9 & 0.307 \\
J1910+1256 & \hphantom{0}2.4 & \hphantom{0}8.3 & 0.187 \\
J0030+0451 & \hphantom{0}2.4 & 12.4 & 0.200 \\
  \hline
  \end{tabular}
  \end{center}
  \tablecomments{The ten pulsars that show the strongest evidence for a common-spectrum process include many pulsars with long observational baselines and low timing RMS, as expected.}
  \label{tab:dropout}
\end{table}

\subsection{Characterizing the statistical significance of Hellings--Downs correlations}

Formally, it is the posterior odds ratio itself that relays the data's support for each model. What it does not tell you is how often noise processes alone could manifest an odds ratio as large as the data gives. While arbitrary rules of thumb have been developed to interpret odds ratios \citep[e.g.,][]{jeffreys1998theory,kr95}, this interpretation is highly problem-specific. However, most analysts would agree that ratios $\sim1$ are inconclusive, while very large or small ratios point to a strong preference for either model. 
In classical hypothesis testing, one computes a detection statistic from the data suspected to contain a signal, then compares the value of the statistic with its background distribution, computed over a population of data sets known to host no signal, and thus representing the null hypothesis. The percentile of the observed detection statistic within the background distribution is known as the $p$-value; it quantifies how incompatible the data are with the null hypothesis (but \emph{not} the probability that the hypothesis of interest is true).

The problem for GW detectors is that it is not possible to construct the background distribution by physically turning off sensitivity to GWs. However, one can operate on the data. For the coincident detection of transient GW signals with ground-based observatories, the null model is realized by applying relative time shifts to the time series of detection statistics from multiple detectors, thus removing the very possibility of coincidence. 
Similar techniques can be applied to the detection of HD correlations in PTA data sets.

Several methods have been developed to perform a frequentist study of the null hypothesis distribution in PTAs \citep{cs16,tlb+17}; the relevant null hypothesis is that of a red process with identical spectral properties in all pulsars, but without any GW-induced inter-pulsar correlations (our so-called common red process). By performing repeated trials of spatial-correlation template scrambles (``sky scrambles'') and Fourier-basis phase offsets (``phase shifts''), we can effectively null any spatial correlations in the true data set, and construct a distribution of our detection statistic (whether frequentist S/N or Bayesian odds ratio) under the null hypothesis. It is with these null distributions that we obtain the $p$-value of our measured statistic.

\begin{figure}[t]
\begin{center}
    \includegraphics[width=0.9\columnwidth]{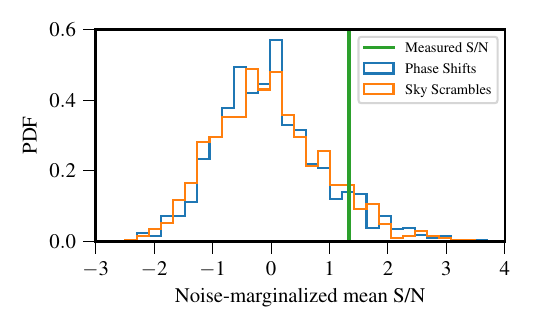}
\end{center}
    \caption{Distribution of the noise-marginalized optimal statistic mean S/N for 1000 phase shifts (blue curve) and 1000 sky scrambles (orange curve). The vertical green line marks the mean S/N measured in the unperturbed model. Higher mean values of the S/N are obtained in 91 phase shifts ($p = 0.091$) and 82 sky scrambles ($p = 0.082$). 
    \label{fig:optstat_shift_scramble}}
\end{figure}

\begin{figure}[t]
\begin{center}
    \includegraphics[width=0.9\columnwidth]{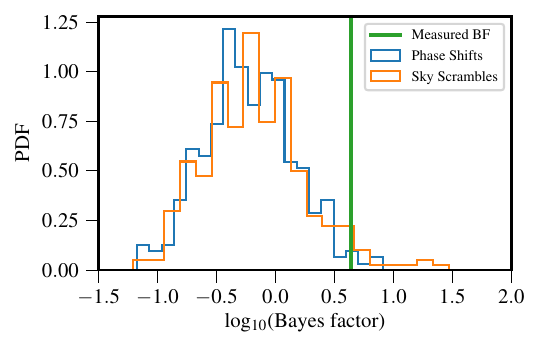}
\end{center}
    \caption{Distribution of correlated-vs-uncorrelated common-process Bayes factor for 300 phase shifts (blue curve) and 300 sky scrambles (orange curve). The vertical green line marks the Bayes factor computed in the unperturbed model. Higher Bayes ratios are obtained in 4 phase shifts ($p = 0.013$) and 13 sky scrambles ($p = 0.043$). The small numbers indicated that statistical error may be large in the $p$-value estimates. 
    \label{fig:bayes_skyscrambles_phaseshifts}}
\end{figure}

In a \emph{phase-shift} analysis, random phase shifts are inserted in the Fourier basis components that describe the GWB process in each pulsar, thus breaking any inter-pulsar correlations that may be present in the data \citep{tlb+17}. Detection statistics are then computed using both frequentist (i.e., the noise-marginalized mean-S/N optimal statistic) and Bayesian (i.e., the Bayes factor for a HD correlated model vs.\ a common-spectrum but spatially uncorrelated model) analyses from 1000 and 300 realizations (respectively) of the phase shifts. The resulting distributions are shown in \autoref{fig:optstat_shift_scramble} and \autoref{fig:bayes_skyscrambles_phaseshifts}. The $p$-values (in this case, the fraction of background samples with statistic higher than observed for the undisturbed model) are $0.091$ and $0.013$.

In a \emph{sky-scramble} analysis, the positions of the pulsars used to compute the expected HD correlations are randomized \citep{cs16, tlb+17}, under the requirement that the scrambled ORF have minimal similarity to the true function.\footnote{Specifically, we measure the match statistic $\bar{M}$ between the ORFs $\Gamma_{ab}$ and $\Gamma'_{ab}$ \citep{tlb+17}:
\begin{equation}
	\bar{M} = \frac{\sum_{a,b \neq a} \Gamma_{ab} \Gamma'_{ab}}{\sqrt{\left(\sum_{a,b \neq a} \Gamma_{ab} \Gamma_{ab} \right) \left(\sum_{a,b \neq a} \Gamma'_{ab} \Gamma'_{ab} \right)}} \,,
\end{equation}
where $a$ and $b$ index the array pulsars, and require that $\bar{M} < 0.1$.} 
Again we compute both frequentist and Bayesian HD detection statistics over large sets of realizations: the resulting background distributions are shown in \autoref{fig:optstat_shift_scramble} and \autoref{fig:bayes_skyscrambles_phaseshifts}.
The optimal-statistic $p$-value agrees closely with its phase-shift counterpart; the Bayes factor $p$ value is higher, but small-number error is likely to be significant.

All of these $p$-values hover around $5\%$, which is much higher than the 3-$\sigma$ (``evidence'') and 5-$\sigma$ (``discovery'') standards of particle physics, corresponding to $p = 0.001$ and $3 \times 10^{-7}$, respectively.
% \footnote{Although one may argue that there is no ``look elsewhere'' effect in our case, so a 3-$\sigma$ finding may already be compelling. Of course, the determination of statistical significance requires much more nuanced considerations \citep{ASA2016}.}
Nevertheless, progressively smaller $p$-values for future data sets would indicate that compelling evidence is accumulating.

\section{Discussion}
\label{sec:discussion}
% !TEX root = nanograv_12p5yr_gwb.tex

As reported in Sec.\ \ref{sec:model_compare}, the \Acp\ posterior has significant support above the upper limits reported in our GWB searches in the $11$-year and $9$-year data sets \citepalias{abb+16, abb+18b}; in fact, almost the entire posterior sits above the most stringent upper limit in the literature ($\Agw < 1 \times 10^{-15}$, \citealt{srl+15}). %However, a revised upper limit from \citet{srl+15}, which will account for solar-system effects using \bayesephem\, is in preparation using the most recent IPTA data release \citep{pdd+19}. 
Without a re-analysis of the data presented in \citet{srl+15}, which is beyond the scope of this work, we cannot fully explain the discrepancy between the results presented in this paper and the upper limit quoted in \citet{srl+15}. A revised analysis of the PPTA data is planned as a part of an upcoming IPTA publication using the DR2 combined data set \citep{pdd+19}; preliminary results show broad consistency with this work. However, we note that the \citeauthor{srl+15} constraint relies on four pulsars, whereas at least ten pulsars in the \citetalias{aab+20} data support a common-spectrum process (see Fig. \ref{fig:dropout_combined}); furthermore, the \citeauthor{srl+15} analysis adopts the DE421 SSE, which even with NANOGrav data yields a lower upper limit than later SSEs \citep{abb+18b,vts+20}. 

In Sec.\ \ref{sec:dataset_compare}, we discuss in detail the discrepancy between the published \citetalias{abb+18b} results and those reported in this paper and find an explanation in the choice of Bayesian prior for the amplitude $A_\mathrm{red}$ of pulsar-intrinsic red-noise processes \citep{hazboun:2020b}. 
While we focus our discussion solely on NANOGrav's previous GWB analyses, we expect the conclusions to apply broadly to all pulsar timing data sets and analyses. 
While the GWB attribution of the common-spectrum process remains inconclusive, in Sec.\ \ref{sec:astro} we consider the broad astrophysical implications of a GWB at the levels encompassed by the \Acp\ posterior.
In Sec.\ \ref{sec:future} we describe the next steps for NANOGrav GWB searches, as well as our expectations for the growth of spatial correlations in future data sets.

\subsection{Comparison of 11-year and 12.5-year results} \label{sec:dataset_compare}

\begin{figure}[t]
\begin{center}
    \includegraphics[width=0.9\columnwidth]{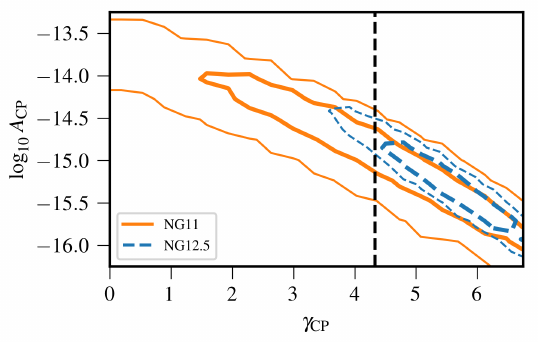}
\end{center}
    \caption{Common-spectrum process parameter posteriors for the \citetalias{aab+20} (dashed curves) and \citetalias{abb+18b} (solid) datasets, as estimated with a five-frequency power-law model under DE438. For each dataset, the two curves trace 1-$\sigma$ and 2-$\sigma$ contours, which appear entirely consistent. The dashed vertical line marks $\gamma = 13/3$, as expected for GWB from SMBHBs.
    \label{fig:11to12.5}}
\end{figure}

We recognize that the common-spectrum amplitude estimated from the $12.5$-year data set ($1.4\mbox{--}2.7 \times 10^{-15}$) may seem surprising when compared to the Bayesian upper limits quoted from analyses of earlier data ($1.45 \times 10^{-15}$ in \citetalias{abb+18b} and $1.5 \times 10^{-15}$ in \citetalias{abb+16}). First, we note that applying the fiducial analysis of this paper (common-spectrum uncorrelated process under DE438) to the $11$-yr dataset results in $A_\mathrm{CP}$-$\gamma_\mathrm{CP}$ posteriors that are entirely consistent with those reported here, as shown in Fig.\ \ref{fig:11to12.5}.

The remaining dissonance between earlier upper limits and the findings of this paper is explained by examining the structure of our analysis. The strength of the Bayesian approach to PTA searches is that it allows for simultaneous modeling of multiple time-correlated processes present in the data. Within the construction of our analysis, amplitude estimates for one such process are sensitive to the priors assumed for the others, especially when the process of interest is still below the threshold of positive detection.

Looking at the $11$-year upper limit specifically (which was quoted as $1.34 \times 10^{-15}$ for a spatially uncorrelated common-spectrum process in \citetalias{abb+18b}), we note that introducing \bayesephem\ corrections with unconstrained priors on Jupiter's orbital perturbation parameters would have necessarily absorbed power from a common-spectrum process, if such a process was present.
Correspondingly, the $11$-year upper limit rises to $1.94 \times 10^{-15}$ if we take DE438 as the fiducial SSE, without corrections \citep{vts+20}.

Even more important, the Bayesian upper limits in \citetalias{abb+16} and \citetalias{abb+18b} were computed by placing a uniform prior on the amplitude of pulsar-intrinsic red noise, which amounts to assuming that loud intrinsic noise is typical among PTA pulsars, rather than exceptional, as suggested by the estimates in this paper. Doing so is conservative with respect to detecting a GWB, but it has the effect of depressing upper limits. As discussed in \cite{hazboun:2020b}, simulations show that injecting a common-spectrum stochastic signal
in synthetic data sets leads to $95\%$ upper limits \emph{lower} than $A^\mathrm{inj}_\mathrm{GWB}$ in $50\%$ of data realizations, \emph{if} intrinsic red noise is given a uniform amplitude prior.

Reweighting the $11$-year upper limit with a log-uniform prior on intrinsic-noise amplitudes yields $2.4 \times 10^{-15}$ under DE438, and $2.1 \times 10^{-15}$ with \bayesephem. Both values are more consistent with the findings of this paper. The differences in data reduction and in the treatment of white noise between $11$-year and $12.5$-year data sets (discussed in Sec.\ \ref{sec:timing_noise}) seem to account for the remaining distance, but those differences are very challenging to evaluate formally, so we do not address them further here.

Altogether, this discussion suggests that past Bayesian upper limits from PTAs may have been overinterpreted in astrophysical terms. Those limits were indeed correct within the Bayesian logic, but they were necessarily affected by our uncertain assumptions. If future data sets bring about a confident GWB detection, our astrophysical conclusions will finally rest on a much stronger basis.
% Would be fun, but not for here, to do a retrospective analysis of LIGO upper limits before and after detection.

\subsection{Astrophysical Implications} \label{sec:astro}

The first hint of a signal from our analysis of \citetalias{aab+20} is indeed tantalizing. However, without definite evidence for HD correlations in the recovered common-spectrum process, there is little we can say about the physical origin of this signal. 
Models have been proposed which give rise to a GWB in the nanohertz frequency range ($\sim 1\mbox{--}100$ nHz) through either primordial GWs from inflation \citep{g75,lms+16}, bursts from networks of cosmic strings \citep{smc07,bos18}, or the mergers of SMBHBs \citep{rr95, p01, jb03, wl03}. 
Black hole mergers are likely the most studied source, though what fraction (if any) of galaxy mergers are able to produce coalescing SMBHBs is virtually unconstrained.  If the common-spectrum process is due to SMBHBs, it would be the first definitive demonstration that SMBHBs are able to form, reach sub-parsec separations, and eventually coalesce due to GW emission.

While the recovered amplitude for the common-spectrum process in this data set is larger than the upper limit on a stochastic GWB quoted in \citetalias{abb+18b}, the qualitative astrophysical conclusions reported there apply to this data set as well (see Sec.\ 5 of \citetalias{abb+18b}). We note also that the amplitude posteriors found here can accommodate many GWB models and assumptions (such as the Kormendy \& Ho measurement of the $M_\mathrm{BH}$--$M_\mathrm{bulge}$ relationship) that had previously been in tension with PTA upper limits. 

The cosmic history of SMBHB mergers is encoded in the shape and amplitude of the GWB strain spectrum they produce \citep{s13c, mop14,rws+14,scm2015,mdf+16,tss17,KelleyEtAl:2017,smc+17,csc19,m19}. %For binaries to reach the PTA band, environmental interactions such as dynamical friction and stellar scattering \citep{bbr80} are required to harden the binary system, and thus the detection of a SMBHB GWB would show that some systems are able to overcome the ``final-parsec problem'' \citep{Yu02,mm03} on a reasonable cosmological timescale. 
At the lower end of the nHz band, signs of the binary hardening mechanism may still be present, %If stellar scattering \citep{q96, mm03} is much more effective than GW radiation, then fewer binaries and thus less GW power will be emitted compared to the pure power-law model from GW emission alone. A circumbinary gas disk can also torque the binary, removing additional energy and angular momentum \citep[][cf.~\citealt{Munoz+2019}]{Ivanov+1999, Cuadra+2009, ks11, Roedig+2012}.  In addition to environmental processes, eccentric binary systems radiate away energy at higher harmonics moving GW energy from lower to higher frequencies~\citep{en07,hmgt15}.  
and we refer the reader to Section 5 of \citetalias{abb+18b} and references therein for further details. % on the information encoded in the low-frequency GWB turnover.
The overall amplitude of the GWB spectrum is determined not only by the number of binaries able to reach the relevant orbital frequencies, but also their distribution of masses \citep{sbs16}.
The GWB amplitude is relatively insensitive to the redshift distribution of sources \citep{p01} except at the highest frequencies, which are affected even more by the local number density and eccentricity distribution of sources \citep{s13c, KelleyEtAl:2017}.
Additionally, the amplitude recovered in this paper, if assumed to be primarily due to a GWB, may imply that the black hole mass function is underestimated, specifically when extrapolated from observations of the local supermassive black hole population \citep{zct19}.

Last, beyond the marginal evidence for HD correlations, we find a broad posterior for the spectral slope $\gamma$ of the common-spectrum process when we allow $\gamma$ to vary.
Therefore, the emerging signal could also be attributed to one of the other cosmological sources capable of producing a nHz GWB. 
The predicted spectral index for these is only slightly different from SMBHBs value of $13/3 ~(\approx 4.33)$: it is $5$ for a primordial GWB~\citep{g05} and $16/3 ~(\approx 5.33)$ for cosmic strings \citep{oms10}.
Data sets with longer timespans and more pulsars will allow for precise parameter estimation in addition to providing confidence toward or against GWB detection.

\subsection{Expectations for the Future} \label{sec:future}

The analysis of NANOGrav pulsar timing data presented in this paper is the first PTA search to show definite evidence for a common-spectrum stochastic signal across an array of pulsars.
However, evidence for the tell-tale quadrupolar HD-correlations is currently lacking, and there are other potential contributors to a common-spectrum process. 
A majority of the pulsars with long observational baselines show the strongest evidence for a common-spectrum process; this subset of pulsars could be starting to show similar spin noise with a consistent spectral index. However, it is unlikely that strong spin noise would appear at a similar amplitude in all millisecond pulsars \citep{lcc+2017}. 
Additionally, the per-pulsar evidence is significantly reduced when we apply \bayesephem, as expected; there remain other solar system effects for which we do not directly account, such as planetary Shapiro delay \citep{tempo2}, that could contribute to the common-spectrum process. 
Finally, there are other sources of systematic noise that we may have uncovered \citep{thk+2016}, and the further potential for sources yet to be diagnosed, all of which would require further study to isolate. 
Thus, attributing the signal uncovered in this work to an astrophysical GWB will necessitate verification with independent pipelines on larger (and/or independent) data sets.

One avenue to validate the processing of timing observations will be the analysis of the ``wideband" version of NANOGrav's $12.5$-year data set, which is produced by a significantly different reduction pipeline \citep{12yr_wideband}. A preliminary analysis of wideband data using the techniques of this paper shows results consistent with those detailed here.
Additionally, our treatment and understanding of pulsar-intrinsic noise will be enhanced soon with the adoption of advanced noise models tailored to each pulsar \citep{simon20}, which include more powerful descriptions of dispersion-measure oscillations among other enhancements.

In the medium term, NANOGrav is compiling its next data set, which adds multiple years of observations and many new pulsars to \citetalias{aab+20}, some of which will have baselines long enough to be incorporated in GW searches. 
If we assume optimistically that the common-spectrum signal identified here is indeed astrophysical, the optimal statistic S/N should then grow by a factor of a few \citep{ptk+20}. %(according to the scaling laws for the intermediate signal regime described in \citet{sejr13}).

Finally, data from the other PTA collaborations will play an important role: the second IPTA data release \citep{pdd+19} includes the $9$-year NANOGrav data set alongside EPTA and PPTA timing observations. The analysis of this joint data set is ongoing, and early results are again consistent with those discussed here. 
Thus, future data sets will be strong arbiters of the astrophysical interpretation of our findings.

NANOGrav's pursuit of a stochastic GWB detection has hardly been linear.
In \citetalias{abb+18b}, we reanalyzed the $9$-year data set using \bayesephem\ and updated the results reported in \citetalias{abb+16} to reflect our new understanding of ephemeris errors.
In this work, we reweighted the $11$-year analysis to account for the emerging physical picture of PTA data quality.
While we cannot foresee how we will revise this $12.5$-year analysis in light of the $15$-year data set, the ouroboric nature of hierarchical Bayesian inference will undoubtedly require some refinements.
The LIGO--Virgo discovery of high-frequency, transient GWs from stellar black-hole binaries appeared meteorically, with incontrovertible statistical significance.  By contrast, the PTA discovery of very-low-frequency %, continuous 
GWs from SMBHBs will emerge from the gradual and not always monotonic accumulation of evidence and arguments.
Still, our GW vista on the unseen universe continues to get brighter.

\acknowledgments

\emph{Author contributions.}
An alphabetical-order author list was used for this paper in recognition of the fact that a large, decade timescale project such as NANOGrav is necessarily the result of the work of many people. All authors contributed to the activities of the NANOGrav collaboration leading to the work presented here, and reviewed the manuscript, text, and figures prior to the paper's submission. 
Additional specific contributions to this paper are as follows.
% 12.5yr Timers
ZA, HB, PRB, HTC, MED, PBD, TD, JAE, RDF, ECF, EF, NG-D, PAG, DCG, MLJ, MTL, DRL, RSL, JL, MAM, CN, DJN, TTP, NSP, SMR, KS, IHS, RS, JKS, RS and SJV developed the 12.5-year data set through a combination of observations, arrival time calculations, data checks and refinements, and timing model development and analysis; additional specific contributions to the data set are summarized in \citetalias{aab+20}.
% Specific individual contributions
JS coordinated the writing of the paper and led the search.
PTB, PRB, SC, JAE, JSH, AMH, KI, ARK, NL, NSP, JS, KS, JPS, SRT, JET, SJV and CAW performed different analyses associated with this work, including exploratory analyses on preliminary versions of the data set. 
SC, JSH, NL, JS, JPS, XS and SRT developed and tested new noise models and created a detailed noise portrait of the 12.5-year data set.
NSP ran the injection analysis in the 11-year data set.
JSH performed the noise parameter comparison between the 11-year data set and the 11-year ``slice" of the 12.5-year data set, and produced the \textit{hasasia} calculations.
SRT developed and executed new Bayesian schemes to assess spatial correlations.
JSH, SRT, MV, and SJV developed and performed new tests of the statistical significance of the common spectrum process.
NJC, XS and MV provided feedback on searches and new analysis techniques.
LZK, CMFM and JS developed the astrophysical interpretation.
JSH, NSP, JS, SRT, MV and SJV prepared the figures and tables.
JSH, LZK, CMFM, NSP, JS, SRT, MV and SJV wrote the paper and collected the bibliography.

\emph{Acknowledgments.}
This work has been carried out by the NANOGrav collaboration, which is part of the International Pulsar Timing Array. 
We thank the members of the IPTA Steering Committee whose comments helped improve and clarify the manuscript. 
We also thank the anonymous reviewers for useful suggestions and comments, which improved the quality of the manuscript. 
The NANOGrav project receives support from National Science Foundation (NSF) Physics Frontiers Center award number 1430284. 
The Arecibo Observatory is a facility of the NSF operated under cooperative agreement (\#AST-1744119) by the University of Central Florida (UCF) in alliance with Universidad Ana G. M\'{e}ndez (UAGM) and Yang Enterprises (YEI), Inc. 
The Green Bank Observatory is a facility of the NSF operated under cooperative agreement by Associated Universities, Inc. The National Radio Astronomy Observatory is a facility of the NSF operated under cooperative agreement by Associated Universities, Inc. 
A majority of the computational work was performed on the Nemo cluster at UWM supported by NSF grant No. 0923409. 
This work made use of the Super Computing System (Spruce Knob) at WVU, which are funded in part by the National Science Foundation EPSCoR Research Infrastructure Improvement Cooperative Agreement \#1003907, the state of West Virginia (WVEPSCoR via the Higher Education Policy Commission) and WVU. 
Part of this research was carried out at the Jet Propulsion Laboratory, California Institute of Technology, under a contract with the National Aeronautics and Space Administration. 
Portions of this work performed at NRL were supported by Office of Naval Research 6.1 funding. 
The Flatiron Institute is supported by the Simons Foundation. 
Pulsar research at UBC is supported by an NSERC Discovery Grant and by the Canadian Institute for Advanced Research. 
JS and MV acknowledge support from the JPL RTD program. 
SBS acknowledges support for this work from NSF grants \#1458952 and \#1815664. SBS is a CIFAR Azrieli Global Scholar in the Gravity and the Extreme Universe program. 
TTP acknowledges support from the MTA-ELTE Extragalactic Astrophysics Research Group, funded by the Hungarian Academy of Sciences (Magyar Tudom\'{a}nyos Akad\'{e}mia), that was used during the development of this research. 

\facilities{Arecibo, GBT}

\software{\texttt{ENTERPRISE} \citep{enterprise}, \texttt{enterprise\_extensions} \citep{enterprise_ext}, \texttt{HASASIA} \citep{hasasia}, \texttt{libstempo} \citep{libstempo}, \texttt{matplotlib} \citep{matplotlib}, \texttt{PTMCMC} \citep{ptmcmc}, \texttt{tempo} \citep{tempo}, \texttt{tempo2} \citep{tempo2}, \texttt{PINT} \citep{pint}} 

%\clearpage

% !TEX root = nanograv_12p5yr_gwb.tex

\appendix

\section{Injection analysis of the NANOGrav 11-year data set} \label{app:inject}

\begin{figure}[t]
	\begin{center}
		\includegraphics[width=\columnwidth]{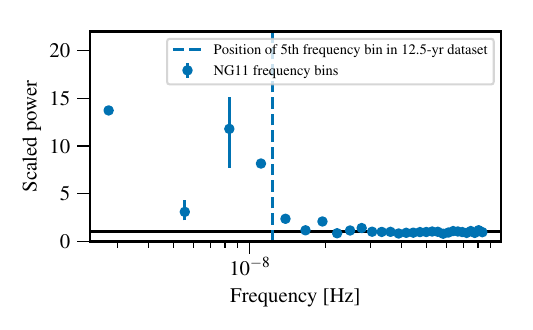}
	\end{center}
	\caption{Response of a common free spectral model's red-noise Fourier-domain components to a GWB injected in the 11-year data set \citetalias{abb+18a}. We plot the component frequency along the horizontal axis, and the ratios of mean estimated component power between injection amplitudes $A_\mathrm{CP} = 5 \times 10^{-15}$ and $A_\mathrm{CP} = 10^{-16}$ along the vertical. Clearly the response to an increasing GWB amplitude is limited to the first few bins. See \autoref{app:inject} for more details.
	\label{fig:largest_inj_power_ratio}}
\end{figure}

To test the response of our real data sets to the presence of a stochastic GWB, we inject a range of GWB amplitudes directly into the $11$-year data set \citepalias{abb+18a}. 
We use the $11$-year data set rather than the current $12.5$-year data set because it does not contain any significant common-spectrum processes and so the GWB injection is able to be cleanly recovered. 
While retaining the TOAs and their corresponding errors from \citetalias{abb+18a}, we inject a stochastic GWB \citep{ccs+2015} using functionality in the {\sc libstempo} software package. Using a power-law model with a spectral index of $\alpha = -2/3$ (i.e., $\gamma = 13/3$), we create ten data set realizations for each characteristic strain amplitude in the range $10^{-16} \leq A_\mathrm{GWB} \leq 5 \times 10^{-15}$. We analyze all realizations with our full detection pipeline. While the complete results of this analysis will be reported in an upcoming publication, here we concentrate on the spectral response of \citetalias{abb+18a} to the presence of the stochastic GWB.

As stated in Sec.~\ref{sec:gw_spectrum}, we calculate the power in each frequency bin using the free spectrum model (see Sec.~\ref{sec:ptalike}) without including HD correlations or \bayesephem.
In \autoref{fig:largest_inj_power_ratio}, we show the ratio of power recovered by each frequency bin between an injection of $A = 5 \times 10^{-15}$ and $A = 1 \times 10^{-16}$. As we can see, the lowest four frequency bins are the most responsive to the presence of a power law GWB in the data set.

\begin{figure}[t]
	\begin{center}
		\includegraphics[width=\columnwidth]{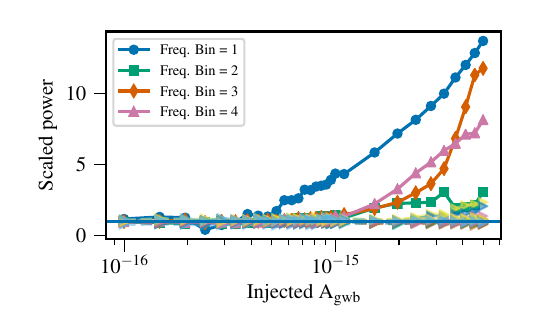}
	\end{center}
	\caption{Response of each frequency from a common free spectral model to the presence of an injected GWB into the 11-year data set (\citepalias{abb+18a}) as a function of the injected GWB amplitude. The x-axis shows the injected GWB amplitude, while the y-axis shows the mean ratio across four realizations of the GWB of the average power in each frequency bin scaled to the mean power in that bin at an injected amplitude of $A = 10^{-16}$. The lowest frequency bin responds to the GWB at much smaller injected amplitudes than the other bins, while the lowest four frequency bins have the strongest response to the presence of the injected GWB at larger amplitudes.}
	\label{fig:ind_freq_bin_power}
\end{figure}

We also examine the evolution of the power in each frequency bin as a function of the injected amplitude. \autoref{fig:ind_freq_bin_power} shows the evolution of the power in each frequency bin, which is scaled to the power in that bin at an injected amplitude of $A = 10^{-16}$.
Due to its power-law nature, the GWB affects the lowest frequency bin at amplitudes much smaller than that for the higher frequency bins. We see again that the lowest four frequency bins are the ones that are most reactive to the presence of a GWB in the data set. This result provides further confirmation that using the five lowest frequencies is sufficient to recover a GWB in the $12.5$-year dataset (Sec.~\ref{sec:gw_spectrum}).

\section{Bayesian Methods} \label{app:methods}

We used Markov chain Monte Carlo (MCMC) methods to stochastically sample the joint posterior of our model parameter spaces, and use Monte Carlo integration to deduce marginalized distributions, where $\int f(\theta)p(\theta|d)d\theta \approx \langle f(\theta_i)\rangle$ for the integral of an arbitrary function $f(\theta)$ over the posterior $p(\theta|d)$ of which the samples $\{\theta_i\}$ are randomly drawn.
Where necessary, we estimated the uncertainty on the marginalized posterior value to be the Monte Carlo sampling error of the location $\hat\theta_x$ of the $x$-th quantile:
\begin{equation}
	\frac{\sqrt{x(1-x)/N}}{p(\theta=\hat\theta_x|d)},
\end{equation}
where $N$ is the number of (quasi-)independent samples in our MCMC chain \citep{wilcox2012introduction}.

As described in \citetalias{abb+18b}, we employ two techniques for model selection based on the relationship between the competing models. For \textit{nested} models that compare the additional presence of a signal to that of noise alone, we used the \textit{Savage-Dickey approximation} \citep{d71}. This requires adequate sampling coverage of low amplitude posterior regions in order to compute the Savage-Dickey density ratio, which corresponds to the prior to posterior density at zero amplitude: Bayes factor $ = p(A=0) / p(A=0|d)$. In practice this means that the method is only useful for moderate model odds contrasts, and while this was used extensively in \citetalias{abb+18b}, the strength of the recovered signal in this paper exceeds the reliability of the Savage-Dickey approximation without additional sampling strategies to explore the low amplitude posterior region. For \textit{disjoint} models, models that are not easily distinguished parametrically, and indeed all model selection in this paper, we used the \textit{product-space method} \citep{cc95,g01,hee15,tvs20}. This recasts model selection as a parameter estimation problem, introducing a model indexing variable that is sampled along with the parameters of the competing models, and which controls which model likelihood is active at each MCMC iteration. The ratio of samples spent in each bin of the model indexing variable returns the posterior odds ratio between models. The efficiency of model transitions is controlled by our prior model probabilities, which we usually set to be equal. However, one can improve the odds ratio computation by performing a pilot run, whose odds ratio estimate can be used to re-weight the models in a follow-up run. This will ensure more equitable chain visitation to each model, after which the model index posterior is re-weighted back to the true model contrast. 

\section{Software} \label{app:software}

% !TEX root = nanograv_12p5yr_gwb.tex

\begin{table*}[ht]
\begin{center}
\scriptsize
\caption{Prior distributions used in all analyses performed in this paper.}
\label{tab:priors}
\begin{tabular}{llll}
\hline\hline
parameter & description & prior & comments \\
\hline

\multicolumn{4}{c}{White Noise} \\[1pt]
$E_{k}$ & EFAC per backend/receiver system & Uniform $[0, 10]$ & single-pulsar analysis only \\
$Q_{k}$ [s] & EQUAD per backend/receiver system & log-Uniform $[-8.5, -5]$ & single-pulsar analysis only \\
$J_{k}$ [s] & ECORR per backend/receiver system & log-Uniform $[-8.5, -5]$ & single-pulsar analysis only \\
\hline

\multicolumn{4}{c}{Red Noise} \\[1pt]
$A_{\rm red}$ & log-Uniform $[-20, -11]$ & one parameter per pulsar  \\
$\gamma_{\rm red}$ & red-noise power-law spectral index & Uniform $[0, 7]$ & one parameter per pulsar \\
\hline

\multicolumn{4}{c}{common process, free spectrum} \\[1pt]
$\rho_{i}$ [s$^{2}$] & power-spectrum coefficients at $f=i/T$ & uniform in $\rho_{i}^{1/2}$ $[10^{-18},10^{-8}]^{a}$ & one parameter per frequency\\
\hline

\multicolumn{4}{c}{common process, broken--power-law spectrum} \\[1pt]
\Acp & broken power-law amplitude & log-Uniform $[-18, -14]$ ($\gamma_\mathrm{CP}=13/3$) & one parameter for PTA \\
& & log-Uniform $[-18, -11]$ ($\gamma_\mathrm{CP}$ varied) & one parameter for PTA \\
$\gamma_{\rm CP}$ & broken--power-law low-freq.\ spectral index & delta function ($\gamma_\mathrm{common}=13/3$) & fixed \\
 & & Uniform $[0,7]$ & one parameter per PTA \\
$\delta$ & broken--power-law high-freq.\ spectral index & delta function ($\delta=0$) & fixed \\
$f_{\rm bend}$ [Hz] & broken--power-law bend frequency & log-Uniform  [$-8.7$,$-7$] & one parameter for PTA \\
\hline

\multicolumn{4}{c}{common process, power-law spectrum} \\[1pt]
\Acp & common process strain amplitude & log-Uniform $[-18, -14]$ ($\gamma_\mathrm{CP}=13/3$) & one parameter for PTA \\
& & log-Uniform $[-18, -11]$ ($\gamma_\mathrm{CP}$ varied) & one parameter for PTA \\
$\gamma_{\rm CP}$ & common process power-law spectral index & delta function ($\gamma_\mathrm{CP}=13/3$)& fixed \\
& & Uniform $[0,7]$ & one parameter for PTA \\
\hline

\multicolumn{4}{c}{\textsc{BayesEphem}} \\[1pt]
$z_{\rm drift}$ [rad/yr] & drift-rate of Earth's orbit about ecliptic $z$-axis & Uniform [$-10^{-9}, 10^{-9}$] & one parameter for PTA \\
$\Delta M_{\rm jupiter}$ [$M_{\odot}$] & perturbation to Jupiter's mass & $\mathcal{N}(0, 1.55\times 10^{-11})$  & one parameter for PTA \\
$\Delta M_{\rm saturn}$ [$M_{\odot}$] & perturbation to Saturn's mass & $\mathcal{N}(0, 8.17\times 10^{-12})$  & one parameter for PTA \\
$\Delta M_{\rm uranus}$ [$M_{\odot}$] & perturbation to Uranus' mass & $\mathcal{N}(0, 5.72\times 10^{-11})$  & one parameter for PTA \\
$\Delta M_{\rm neptune}$ [$M_{\odot}$] & perturbation to Neptune's mass & $\mathcal{N}(0, 7.96\times 10^{-11})$  & one parameter for PTA \\
PCA$_{i}$ & $i$th PCA component of Jupiter's orbit & Uniform $[-0.05, 0.05]$ & six parameters for PTA \\
\hline

\end{tabular}
\end{center}
\end{table*}

We used the software packages \texttt{enterprise} \citep{enterprise} and \texttt{enterprise\_extensions} \citep{enterprise_ext} to perform the Bayesian and frequentist searches. These packages implement the signal models, likelihood, and priors. Our Bayesian priors for all parameters are described in \autoref{tab:priors}. 
We used the software package \texttt{PTMCMCSampler} \citep{ptmcmc} to perform the MCMC for the Bayesian searches. 
We primarily used adaptive Metropolis and differential evolution jump proposals. 
For some analyses, we used draws from empirical distributions to sample the pulsars' red noise parameters, 
with the empirical distributions constructed from posteriors obtained from previous Bayesian analyses. 
These draws significantly decreased the number of samples needed for the pulsars' red noise parameters 
to burn in. This technique was first used to analyze the $11$-year data set for GWs from individual 
SMBHBs, and a detailed description can be found in Appendix B of \citet{aab+19}.

\clearpage
%\pagebreak

%\bibliographystyle{yahapj}
\bibliography{bib}
%MV: Is there a way to show the abbreviation among the refs?

\end{document}